\documentclass[12pt]{iopart}
\usepackage{graphicx}
\eqnobysec

\begin{document}


\title[Analysis of reflection coefficients]
{Analysis of reflection coefficients
for the Fokker-Planck equation}

\author{Toru Miyazawa}

\address{Department of Physics, Gakushuin University, 
Tokyo 171-8588, Japan}
\ead{toru.miyazawa@gakushuin.ac.jp}
\begin{abstract}
Mathematical structure of the reflection coefficients for the 
one-dimensional Fokker-Planck equation is studied.
A new formalism using differential operators is introduced and applied to the analysis in high- and low-energy regions. 
Formulas for high-energy and low-energy expansions are derived, and expressions for the coefficients of the expansion, as well as the remainder terms, are obtained for general forms of the potential. Conditions for the validity of these expansions are discussed on the basis of the analysis of the remainder terms.
\end{abstract}

\maketitle


\section{Introduction}
It is well known that the steady-state Schr\"odinger equation
\begin{equation}
\label{1eq1.1}
-\frac{d^2}{dx^2}\psi(x)+V_{\rm S}(x)\psi(x)=k^2\psi(x)
\end{equation}
is equivalent to the Fokker-Planck eigenvalue equation \cite{risken}
\begin{equation}
\label{1eq1.2}
-\frac{d^2}{dx^2}\phi(x)+2 \frac{d}{dx}[f(x)\phi(x)]=k^2\phi(x).
\end{equation}
The time-dependent Fokker-Planck equation corresponding to (\ref{1eq1.2}) describes 
diffusion in a potential $V(x)$, where
\begin{equation}
\label{1eq1.3}
f(x)=-\frac{1}{2}\frac{d}{dx}V(x).
\end{equation}
The correspondence between (\ref{1eq1.1}) and (\ref{1eq1.2}) is given by
\begin{equation}
\psi(x)=e^{V(x)/2}\phi(x), \qquad V_{\rm S}(x)=f'(x)+f^2(x).
\end{equation}

We define the transmission and reflection coefficients for a finite interval $(x_1,x_2)$ as follows.
Let $\bar V(x)$ be the function which is identical with $V(x)$ inside the interval $(x_1,x_2)$ 
and constant outside:
\begin{equation}
\label{1eq1.5}
\bar V(x) =
\cases{
V(x_1)
& $x \leq x_1$
\\
V(x)
& $x_1 < x <x_2$
\\
V(x_2)
& $ x_2 \leq x$
}.
\end{equation}
We define $\bar f(x) \equiv -(1/2)(d/dx)\bar V(x)$ just like (\ref{1eq1.3}), and consider equation (\ref{1eq1.2}) with $f(x)$ replaced by $\bar f(x)$. 
(In general, delta functions appear at $x=x_1$ and $x=x_2$ on the left-hand side of this equation.) Since $\bar f(x)=0$ outside $(x_1,x_2)$, this equation has two independent solutions of the form
\refstepcounter{equation}
\label{1eq1.6}
\addtocounter{equation}{-1}
\numparts
\begin{equation}
\phi_1(x)
=
\cases{
e^{ik(x-x_1)} +R_l(x_2,x_1;k)\,e^{-ik(x-x_1)}  & $x<x_1$
\\
e^{-[V(x_2)-V(x_1)]/2} \tau(x_2,x_1;k)\, e^{ik(x-x_2)}  & $x>x_2$
},
\end{equation}
\begin{eqnarray}
\phi_2(x)
=
\cases{
e^{[V(x_2)-V(x_1)]/2} \tau (x_2,x_1;k)\, e^{-ik(x-x_1)} & $x<x_1$
\\
e^{-ik(x-x_2)} +R_r(x_2,x_1;k)\,e^{ik(x-x_2)} & $x>x_2$
}.
\end{eqnarray}
This defines the transmission coefficient $\tau$, the right reflection coefficient $R_r$, 
and the left reflection coefficient $R_l$ for the 
interval $(x_1,x_2)$. 
The factor $e^{\pm [V(x_2)-V(x_1)]/2}$ in front of $\tau$ in the above equations comes from the factor $e^{V(x)/2}$ in the first equation of (1.4).
If we define $\psi_1(x) \equiv e^{[V(x)-V(x_1)]/2} \phi_1(x)$ and $\psi_2(x) \equiv e^{[V(x)-V(x_2)]/2} \phi_2(x)$, then $\psi_1$ and $\psi_2$ are two independent solutions of the corresponding Schr\"odinger equation. Equations (1.6a) and (1.6b) can be rewritten in terms of $\psi_1$ and $\psi_2$ as
\begin{equation}
\psi_1(x)
=
\cases{
e^{ik(x-x_1)} +R_l(x_2,x_1;k)\,e^{-ik(x-x_1)}  & $x<x_1$
\\
\tau(x_2,x_1;k)\, e^{ik(x-x_2)}  & $x>x_2$
},
\end{equation}
\begin{eqnarray}
\psi_2(x)
=
\cases{
\tau (x_2,x_1;k)\, e^{-ik(x-x_1)} & $x<x_1$
\\
e^{-ik(x-x_2)} +R_r(x_2,x_1;k)\,e^{ik(x-x_2)} & $x>x_2$
},
\end{eqnarray}
\endnumparts
in agreement with the standard definition of the transmission and reflection coefficients. 
 Many properties of equation~(\ref{1eq1.2}) or equation (\ref{1eq1.1}) can be known from these scattering coefficients. 

Our object of study in this paper is the reflection coefficients for semi-infinite intervals, 
$R_r(x,-\infty;k)$ and $R_l(\infty,x;k)$, which play particularly important roles in one-dimensional problems.
When considering a problem on the entire line in one dimension, $-\infty<x<+\infty$, we are obliged to deal with semi-infinite intervals.
For example, the Green function is expressed in terms of $R_r(x,-\infty;k)$ and $R_l(\infty,x;k)$.
Let $G_{\rm S}(x,x';k)$ be the Green function for the Schr\"odinger equation (\ref{1eq1.1}), satisfying
\begin{equation}
\left[
\frac{\partial^2}{\partial x^2}-V_{\rm S}(x)+k^2
\right]
G_{\rm S}(x,x';k)=\delta(x-x')
\end{equation}
with the boundary condition $G_{\rm S}(x,x';k+i\epsilon) \to 0$ as $\vert x-x'\vert \to \infty$. (Here $k$ is real and $\epsilon$ is a positive infinitesimal.) 
This Green function can be expressed as\footnote{
This expression, and similar expressions for the Green's function, 
will be discussed in another paper.
}
\begin{equation}
\label{1eq1.8}
\fl
G_{\rm S}(x,x';k)=
\frac{-i}{2k\sqrt{[1-S(x;k)][1-S(x';k)]}}
\exp\left[
ik(x-x')-ik\int_{x'}^x S(z;k)\,dz
\right]
\end{equation}
for $x\geq x'$, where
\begin{equation}
S(x;k) \equiv \frac{R_l(\infty,x;k)}{1+R_l(\infty,x;k)}+\frac{R_r(x,-\infty;k)}{1+R_r(x,-\infty;k)}.
\end{equation}
Therefore, analytic properties of the Green function for the Schr\"odinger equation can be known by studying 
$R_r(x,-\infty;k)$ and $R_l(\infty,x;k)$ for the Fokker-Planck equation.

In this paper we investigate the behavior of these reflection coefficients 
in high-energy (large-$\vert k \vert$) and low-energy (small-$\vert k \vert$) regions.
We shall deal only with $R_r(x,-\infty;k)$ since $R_l$ has the same structure as $R_r$.
We assume that $k$ is, in general, a complex number with ${\rm Im}\,k\geq0$. 

The analysis of scattering coefficients for the Schr\"odinger equation has a very long history [2--4]. Even recently, the high- and low-energy asymptotic expansions of the reflection coefficients and related quantities continue to be studied actively by many researchers [5--10]. 
On the other hand, although the equivalence between the Schr\"odinger equation and the Fokker-Planck equations has been well known for a long time, little attention has been paid to the reflection coefficients for the Fokker-Planck equation. 
Actually, the structure of the reflection coefficients is more transparent for the Fokker-Planck equation than for the Schr\"odinger equation. By dealing with the Fokker-Planck equation rather than the Schr\"odinger equation, we can carry out the analysis in a more systematical way, as we shall see in this paper.
 
Conventional methods used for the Schr\"odinger equation mostly involve estimating a solution of an integral equation. In this paper we take a totally different approach. 
It is a characteristic of the reflection coefficients (and related quantities such as the Weyl $m$-function) that they satisfy a nonlinear differential equation of Riccati type.
In our method, the Riccati equation is transformed into a linear partial differential equation for two variables, and the derivation of the asymptotic expansions is reduced to a manipulation of linear operators. 
In this method, the high-energy expansion and the low-energy expansion can be treated on an equal footing. 

In studying an asymptotic expansion, it is essential to estimate the remainder term. In conventional methods, this procedure often calls for a severe restriction on the potential, requiring it to belong to a certain limited class such as $L^1$, $L^2$, or the Faddeev class. In our method, the remainder term is expressed in a fairly compact form which is valid even if the potential is infinite at $x=\pm \infty$. As a result, this method is applicable to a much larger class of potentials. 

The potential $V(x)$ is a real function of $x$. (In this paper we always use the term ^^ ^^ potential" to mean the 
Fokker-Planck potential $V(x)$, not the Schr\"odinger potential $V_{\rm S}(x)$.)
Since we shall deal only with $R_r(x,-\infty;k)$, the potential need not be defined on the entire line. We assume that $V(x)$ is defined in $-\infty<x<x_{\rm max}$ with some $x_{\rm max}$, and that $V(x)$ takes a finite value for each $x$ in this region. 
(For example, $x_{\rm max}=0$ for $V(x)=\log \vert x \vert$. 
If the potential is defined everywhere, then $x_{\max}=+\infty$.) 

We shall allow $V(x)$ to be either finite or infinite in the limit $x\to -\infty$.
The only requirement we impose on the asymptotic behavior of the potential as $x \to -\infty$ 
is that the function $f(x)$ (defined by (\ref{1eq1.3})) 
should either converge smoothly or diverge smoothly in the following sense:
If $f(-\infty)$ is finite, we assume that all the derivatives of $f(x)$ vanish in the limit $x \to -\infty$,
and that they are all monotone for sufficiently large $(-x)$.
(In fact, this smoothness condition can be relaxed in many cases, but we shall assume this rather strict condition in order to simplify the explanation.)
If $f(-\infty)$ is either $+\infty$ or $-\infty$, then  $1/f(x)$ and all its derivatives
are assumed to vanish as $x \to -\infty$. 
We do not deal with potentials that show oscillatory behavior
at infinity. 
Other conditions on $V(x)$ will be specified when they become necessary. 

In our formalism we deal with the scattering coefficients in a generalized form, which will be defined in the next section. We set up a general framework in section~3, and derive the formulas for low- and high-energy expansions in sections~4 and 5, respectively.


\section{Generalized scattering coefficients}
Let $\xi$ be a real variable, $-1 <\xi < 1$.
We define
\refstepcounter{equation}
\label{1eq4.1}
\addtocounter{equation}{-1}
\numparts
\begin{eqnarray}
\label{1eq4.1a}
{\bar R}_r(x,y;\xi;k)
& \equiv \frac{R_r(x,y;k) -\xi}{1-\xi R_r(x,y;k)},
\\
\label{1eq4.1b}
{\bar R}_l(x,y;\xi;k) 
&\equiv  R_l(x,y;k) + \frac{\xi \tau^2(x,y;k)}{1-\xi R_r(x,y;k)},
 \\
 \label{1eq4.1c}
{\bar \tau}(x,y;\xi;k) 
&\equiv \frac{\sqrt{1-\xi^2}\,\tau(x,y;k)}{1-\xi R_r(x,y;k)}.
\end{eqnarray}
\endnumparts
(See \cite{algebraic} for the background of these definitions\footnote{
In \cite{algebraic}, these quantities  are defined in a more generalized form, with one more additional variable $\xi'$. The $\bar R_r$, $\bar R_l$, and $\bar \tau$ in the present paper correspond to the ones with $\xi'=0$. 
}. 
In fact, they are equivalent to the scattering coefficients for a potential that has a discontinuity at the right endpoint of the interval.)
Since $\tau(x,x;k)=1$ and $R_r(x,x;k)=R_l(x,x;k)=0$, we have
\begin{equation}
\label{6eq4.2}
\fl
\bar \tau(x,x;\xi;k)=\sqrt{1-\xi^2}, \qquad
\bar R_r(x,x;\xi;k)=-\xi, \qquad
\bar R_l(x,x;\xi;k)=\xi.
\end{equation}
 
Sometimes it is convenient to define $W$ by
\begin{equation}
\label{1eq4.4}
\xi \equiv \tanh \frac{W-V(x)}{2}
\qquad {\rm or} \qquad 
W\equiv \log \frac{1+\xi}{1-\xi}+V(x),
\end{equation}
and take $\{x,y,W,k\}$, rather than $\{x,y,\xi,k\}$, as independent variables.
We shall specify the set of independent variables by writing the argument $\xi$ or $W$ explicitly. (We shall often omit to write the argument $k$.)
The original scattering coefficients $\tau,R_r,R_l$ are recovered from $\bar \tau, \bar R_r, \bar R_l$ by setting $\xi=0$ or $W=V(x)$.
For $k=0$, we have~\cite{algebraic}
\refstepcounter{equation}
\label{1eq4.19}
\addtocounter{equation}{-1}
\numparts
\begin{eqnarray}
\label{1eq4.19a}
\bar \tau(x,y;W;k=0)={\rm sech}\,\frac{W-V(y)}{2}, 
\\
\label{1eq4.19b}
\bar R_r(x,y;W;k=0)=-\bar R_l(x,y;W;k=0)=-\tanh \frac{W-V(y)}{2}.
\end{eqnarray}
\endnumparts


\section{Basic formalism}
We consider the set of two-variable functions $g(x,\xi)$ which are defined in $-\infty<x<x_{\rm max}$ and $-1< \xi< 1$, and which are analytic with respect to $\xi$ in this interval.
The generalized reflection coefficient $\bar R_r(x,-\infty;\xi)$ is one of such functions.
From time to time we also regard them as functions of $x$ and $W$, with $W$ defined by (\ref{1eq4.4}). In that case the functions $g(x,W)$ are analytic in $-\infty<W<+\infty$.

Let us define the operators $\cal A$ and $\cal B$
acting on these functions as
\begin{equation}
\label{1eq5.1}
{\cal A}\, g(x,\xi) \equiv \left[\frac{\partial}{\partial x}
+f(x)(1-\xi^2)\frac{\partial}{\partial \xi} \right] g(x,\xi), 
\end{equation}
\begin{equation}
\label{1eq5.2}
{\cal B}\, g(x,\xi) 
\equiv \left(\frac{1+\xi^2}{1-\xi^2}+\xi \frac{\partial}{\partial \xi} \right) g(x,\xi)
=(1-\xi^2) \frac{\partial}{\partial \xi}\frac{\xi}{1-\xi^2}g(x,\xi).
\end{equation}
If we take $\{x,W\}$ as independent variables instead of $\{x,\xi\}$, 
the above definitions read
\begin{equation}
\label{1eq5.3}
{\cal A}\, g(x,W) \equiv \frac{\partial}{\partial x} g(x,W), 
\end{equation}
\begin{equation}
\label{1eq5.4}
{\cal B}\, g(x,W) 
\equiv \left(\cosh[W-V(x)] + \sinh[W-V(x)] \frac{\partial}{\partial W}\right)g(x,W).
\end{equation}
It can be shown that $\bar R_r(x,-\infty;\xi)$ satisfies the partial differential equation~\cite{algebraic}
\begin{equation}
\label{1eq5.5}
({\cal A}- 2ik {\cal B})\bar R_r(x,-\infty;\xi)=4ik \frac{\xi}{1-\xi^2}.
\end{equation}
(There is an algebraic background for this equation; see \cite{algebraic} for details.)
We shall use this equation as a basis for our analysis.

Let $\Omega^{[V]}_k$ denote the set of functions $g(x,\xi)$ which are continuous and piecewise differentiable with respect to $x$, analytic with respect to $\xi$, and which satisfy
\begin{equation}
\label{1eq5.16}
\lim_{z \to -\infty}
\frac{\bar \tau^2(x,z;\xi)}{1-\bar R_l^2(x,z;\xi)}\, 
g(z,\bar R_l(x,z;\xi))=0.
\end{equation}
for any $x$ in $-\infty<x<x_{\rm max}$.
(Whether a given function $g(x,\xi)$ satisfies (\ref{1eq5.16}) or not can be decided by using the asymptotic forms of $\bar \tau$ and $\bar R_l$ shown in appendix~A.)
If we restrict the domain of ${\cal A}-2ik {\cal B}$ to $\Omega^{[V]}_k$, then it has an inverse given by
\begin{equation}
\label{1eq5.13}
\frac{1}{{\cal A}-2ik {\cal B}}\,g(x,\xi)=
\int_{-\infty}^x
\frac{\bar \tau^2(x,z;\xi)}{1-\bar R_l^2(x,z;\xi)}\, g(z,\bar R_l(x,z;\xi)) \,dz.
\end{equation}
(The proof is given in appendix~B.) 
In other words, for any $g(x,\xi)$ belonging to $\Omega^{[V]}_k$, the operator $({\cal A}-2ik {\cal B})^{-1}$ given by (\ref{1eq5.13}) satisfies
\begin{equation}
\label{1eq5.15}
\frac{1}{{\cal A}-2ik {\cal B}}({\cal A}-2ik {\cal B}) g(x,\xi)=g(x,\xi).
\end{equation}

When $\{x,W\}$ are used as independent variables, 
equations (\ref{1eq5.13}) and (\ref{1eq5.16}) read
\begin{equation}
\label{3eq5.36}
\fl
\frac{1}{{\cal A}-2ik {\cal B}}\,g(x,W)=
\int_{-\infty}^x
\frac{\bar \tau^2(x,z;W)}{1-\bar R_l^2(x,z;W)}\, 
g\Bigl(z,V(z)+\log \frac{1+\bar R_l(x,z;W)}{1-\bar R_l(x,z;W)}\Bigr) \,dz,
\end{equation}
and
\begin{equation}
\label{3eq5.37}
\fl
\lim_{z \to -\infty}
\frac{\bar \tau^2(x,z;W)}{1-\bar R_l^2(x,z;W)}\, 
g\Bigl(z,V(z)+\log \frac{1+\bar R_l(x,z;W)}{1-\bar R_l(x,z;W)}\Bigr)=0.
\end{equation}
Condition (\ref{3eq5.37}) takes a  simple form for $k=0$;
substituting (\ref{1eq4.19}) we obtain
\begin{equation}
\label{2eq5.17}
\lim_{x\to -\infty}
g(x,W)=0.
\end{equation}

We may notice that (\ref{2eq5.17}) is not satisfied for $g(x,W)=\bar R_r(x,-\infty;W)$, since
\begin{equation}
\label{4eq5.22}
\bar R_r(x,-\infty;W;k=0)=\tanh\frac{V(-\infty)-W}{2}.
\end{equation}
(See (\ref{1eq4.19b}).) 
If we take $g=\bar R_r+\xi$ instead of $g=\bar R_r$, then (\ref{2eq5.17}) is satisfied. 
(It is obvious that $\xi=\tanh\{[W-V(x)]/2\}$ cancels the right-hand side of (\ref{4eq5.22}) in the limit $x \to -\infty$.)
More generally, we can show that 
\begin{equation}
\label{4eq5.24}
\bar R_r(x,-\infty;\xi;k)+\xi \in \Omega^{[V]}_k
\end{equation}
for any $k$ in the region ${\rm Im}\,k \geq 0$. (See appendix~C for a proof.) 

Using 
$
({\cal A}-2ik{\cal B})\xi =(1-\xi^2)f(x) -4ik\xi/(1-\xi^2)
$,
we rewrite (\ref{1eq5.5}) as
\begin{equation}
\label{1eq5.18}
({\cal A}- 2ik {\cal B})[\bar R_r(x,-\infty;\xi)+\xi]=(1-\xi^2)f(x).
\end{equation}
Since $\bar R_r+\xi \in \Omega^{[V]}_k$, we can apply $({\cal A}- 2ik {\cal B})^{-1}$ 
to both sides of (\ref{1eq5.18}) and obtain
\begin{equation}
\label{1eq5.19}
\bar R_r(x,-\infty;\xi)=-\xi + \frac{1}{{\cal A}-2ik{\cal B}}(1-\xi^2) f(x).
\end{equation}
With (\ref{1eq5.13}), this expression reads
\begin{equation}
\label{1eq4.11}
\bar R_r(x,-\infty;\xi)=-\xi + \int_{-\infty}^x dz f(z) \bar \tau^2(x,z;\xi).
\end{equation}
Equation (\ref{1eq5.19}) is the basic expression for $\bar R_r$. We can derive from it expansions in powers of $k$ and $1/k$ by a simple manipulation of operators, as we shall now see.

From (\ref{1eq5.3}) we can see that the inverse of ${\cal A}$ is given by
\begin{equation}
\label{1eq5.22}
{\cal A}^{-1} g(x,W)=\int_{-\infty}^x g(z,W)\,dz.
\end{equation}
Obviously ${\cal A}^{-1} {\cal A} g=g$ holds provided that $g$ satisfies condition (\ref{2eq5.17}). We can also derive (\ref{1eq5.22}) from (\ref{1eq5.13}) by setting $k=0$ and using (\ref{1eq4.19}).
The inverse of ${\cal B}$ is obtained form the last expression of (\ref{1eq5.2}) as
\begin{equation}
\label{1eq5.25}
{\cal B}^{-1}g(x,\xi)=\frac{1-\xi^2}{\xi}\int_0^\xi\frac{1}{1-\xi^2}\,g(x,\xi)\,d\xi.
\end{equation}
We can easily see that ${\cal B}^{-1}{\cal B}\,g(x,\xi)=g(x,\xi)$ holds as long as
$
\lim_{\xi \to 0} \xi g(x,\xi)=0
$.
Since $g(x,\xi)$ is assumed to be analytic in $-1<\xi<1$, this condition is automatically satisfied. 

Let us define
\begin{equation}
\label{1eq5.27}
{\cal L}\equiv 2 {\cal A}^{-1} {\cal B}, \qquad 
{\cal L}^{-1}=\case{1}{2}{\cal B}^{-1}{\cal A}.
\end{equation}
For an arbitrary positive integer $N$, we can express $({\cal A}-2ik {\cal B})^{-1}$ as
\refstepcounter{equation}
\label{3eq5.35}
\addtocounter{equation}{-1}
\numparts
\begin{eqnarray}
\label{3eq5.35a}
\frac{1}{{\cal A}-2ik {\cal B}}
&=\left[1+ik{\cal L}+(ik)^2{\cal L}^2 + \cdots + (ik)^N {\cal L}^N\right]{\cal A}^{-1}
\nonumber \\
&\qquad +(ik)^{N+1}\frac{1}{{\cal A}-2ik {\cal B}}\,{\cal A}{\cal L}^{N+1}{\cal A}^{-1},
\fl
\end{eqnarray}
and
\begin{eqnarray}
\label{3eq5.35b}
\fl
\frac{1}{{\cal A}-2ik {\cal B}}
&=-\frac{1}{2ik}\left[
1+\frac{1}{ik}{\cal L}^{-1}+\frac{1}{(ik)^2}\left({\cal L}^{-1}\right)^2+\cdots
+\frac{1}{(ik)^{N-1}}\left({\cal L}^{-1}\right)^{N-1}
\right]{\cal B}^{-1}
\nonumber \\
\fl
&\qquad +\frac{1}{(ik)^N}\frac{1}{{\cal A}-2ik {\cal B}}\, {\cal B}\left({\cal L}^{-1}\right)^N{\cal B}^{-1}.
\end{eqnarray}
\endnumparts 
The expansions of $\bar R_r$ are obtained by substituting these expressions into (\ref{1eq5.19}).


\section{Low-energy expansion}
Let us first introduce some notation for integrals that will appear in the expansion. 
We define, for $n=1,2,3,\ldots$ and $-\infty \leq a\leq b \leq \infty$, 
\begin{equation}
\label{1eq6.1}
\fl
[s_1,s_2,\ldots,s_n]_a^b
\equiv 
\int_a^b dz_1 \int_{z_1}^b dz_2 \int_{z_2}^b dz_3 \cdots\int_{z_{n-1}}^b dz_n 
\exp\biggl[\sum_{j=1}^n s_j V(z_j)\biggr],
\end{equation}
where each $s_j$ is either $+1$ or $-1$. 
When $V(-\infty)=V_0 \neq \pm \infty$, we use the notation
\begin{eqnarray}
\label{1eq6.2}
\fl
&(\pm, s_2,s_3, \ldots,s_n]_{-\infty}^x 
\equiv e^{V_0} [-1,s_2,s_3,\ldots,s_n]_{-\infty}^x-e^{-V_0} [+1,s_2,s_3,\ldots,s_n]_{-\infty}^x
\nonumber \\
\fl
& \qquad \qquad\qquad=2\int_{-\infty}^x dz_1 \int_{z_1}^x dz_2  \cdots\int_{z_{n-1}}^x dz_n 
\sinh [V_0-V(z_1)]\exp\biggl[\sum_{j=2}^n s_j V(z_j)\biggr].
\end{eqnarray}

Substituting (\ref{3eq5.35a}) into (\ref{1eq5.19}) yields
\begin{equation}
\label{3eq6.6}
\bar R_r(x,-\infty)
=
\bar r_0 +ik \bar r_1 +(ik)^2 \bar r_2 + \cdots + (ik)^N \bar r_N + \bar \rho_N,
\end{equation}
where
\begin{equation}
\label{3eq6.7}
\bar r_0 \equiv {\cal A}^{-1} (1-\xi^2) f(x) -\xi,
\qquad
\bar r_n \equiv{\cal L}^n (\bar  r_0+ \xi) \quad (n \geq 1),
\end{equation}
\begin{equation}
\label{3eq6.8}
\bar \rho_N \equiv (ik)^{N+1}\frac{1}{{\cal A}-2ik{\cal B}}
\,{\cal A}\,\bar r_{N+1}.
\end{equation}
The first expression of (\ref{3eq6.7}) can be calculated by using (\ref{1eq5.22}) as
\begin{equation}
\label{1eq6.7}
\bar r_0
=\int_{-\infty}^x {\rm sech}^2\frac{W-V(z)}{2}\,f(z)\,dz -\xi
=-\tanh\frac{W-V(-\infty)}{2},
\end{equation}
which agrees with (\ref{4eq5.22}).
This means, according to the behavior of $V(x)$ as $x \to -\infty$,
\begin{eqnarray}
\label{1eq6.11}
\bar r_0
&=\pm 1, \qquad & V(-\infty)=\pm \infty, 
\nonumber \\
&=-\tanh \case{W-V_0}{2}, \qquad & V(-\infty)=V_0.
\end{eqnarray}
From now on, we take $\{x, W\}$ as independent variables. 
The second expression of (\ref{3eq6.7}) is written in terms of $W$ as
\begin{equation}
\label{3eq6.11}
\bar r_n(x,W)={\cal L}^n
\left(\bar r_0+ \tanh\frac{W-V(x)}{2}\right) \qquad (n\geq 1),
\end{equation}
with $\bar r_0$ given by (\ref{1eq6.11}).
As can be seen from (\ref{1eq5.4}) and (\ref{1eq5.22}), the operator ${\cal L}$ acts as
\begin{equation}
\label{1eq6.5}
{\cal L}\,g(x,W)=\int_{-\infty}^x 
\left( e^{-V(z)}\hat{\cal J}_+^{(2)}+ e^{V(z)}\hat{\cal J}_-^{(2)}\right)g(z,W)\,dz,
\end{equation}
where we have defined the operators
\begin{equation}
\label{1eq6.4}
\hat{\cal J}_+^{(2)}\equiv e^W\left(1+\frac{\partial}{\partial W}\right),
\qquad
\hat{\cal J}_-^{(2)} \equiv e^{-W} \left(1-\frac{\partial}{\partial W} \right).
\end{equation}
The right-hand side of (\ref{3eq6.11}) can be calculated by carrying out the integration of (\ref{1eq6.5}) repeatedly. 
The result is expressed in terms of the integrals (\ref{1eq6.1}) and (\ref{1eq6.2}):
\begin{eqnarray}
\label{1eq6.18}
\fl
\bar r_n
=\sum_{\{s_1,\ldots,s_{n-1}\}} C^+_{s_1,s_2,\ldots,s_{n-1}}(W)\, [-1,s_1,s_2,\ldots,s_{n-1}]_{-\infty}^x,
 \qquad
& V(-\infty)=+\infty,
\nonumber \\
\fl \quad
=\sum_{\{s_1,\ldots,s_{n-1}\}} C^-_{s_1,s_2,\ldots,s_{n-1}}(W)\,
[+1,s_1,s_2,\ldots,s_{n-1}]_{-\infty}^x, 
\qquad
& V(-\infty)=-\infty,
\nonumber \\
\fl \quad
=\sum_{\{s_1,\ldots,s_{n-1}\}} D_{s_1,s_2,\ldots,s_{n-1}}(W)\,
(\,\pm,s_1,s_2,\ldots,s_{n-1}]_{-\infty}^x,
\qquad
& V(-\infty)=V_0,
\end{eqnarray}
where
\refstepcounter{equation}
\label{1eq6.19}
\addtocounter{equation}{-1}
\numparts
\begin{eqnarray}
\label{1eq6.19a}
C^+_{s_1,s_2,\ldots,s_{n-1}}(W)
&\equiv 2 \hat{\cal J}^{(2)}_{-s_{n-1}}\cdots
\hat{\cal J}^{(2)}_{-s_2}\hat{\cal J}^{(2)}_{-s_1}\, e^W,
\\
\label{1eq6.19b}
C^-_{s_1,s_2,\ldots,s_{n-1}}(W) 
&\equiv -2 \hat{\cal J}^{(2)}_{-s_{n-1}}\cdots
\hat{\cal J}^{(2)}_{-s_2}\hat{\cal J}^{(2)}_{-s_1}\, e^{-W},
\\
\label{1eq6.19c}
D_{s_1,s_2,\ldots,s_{n-1}}(W) 
&\equiv \case{1}{2} \hat{\cal J}^{(2)}_{-s_{n-1}}\cdots
\hat{\cal J}^{(2)}_{-s_2}\hat{\cal J}^{(2)}_{-s_1}\, {\rm sech}^2\, \case{W-V_0}{2}.
\end{eqnarray}
\endnumparts
The sums in (\ref{1eq6.18}) are over $s_1=\pm1, s_2=\pm1, \ldots, s_{n-1}=\pm1.$
In (\ref{1eq6.19}), the symbol $\hat{\cal J}^{(2)}_{-s_i}$ stands for $\hat{\cal J}^{(2)}_-$ and
 $\hat{\cal J}^{(2)}_+$ for $s_i=+1$ and $s_i=-1$, respectively.
More explicit expressions of $C^\pm$ and $D$ are given in appendix~D.
Using (\ref{3eq5.36}), we can write (\ref{3eq6.8}) as
\begin{equation}
\label{1eq6.16}
\fl
\bar \rho_N =(ik)^{N+1}\int_{-\infty}^x 
\frac{\bar \tau^2(x,z)}{1-\bar R_l^2(x,z)}
\,q_{N+1}\Bigl(z,V(z)+\log \frac{1+\bar R_l(x,z)}{1-\bar R_l(x,z)}\Bigr)\,dz,
\end{equation}
\begin{equation}
\label{1eq6.17}
q_n(x,W) \equiv \frac{\partial}{\partial x} \bar r_n(x,W).
\end{equation}
Explicit expressions of $\bar \rho_n$ for general $n$ are shown in appendix~D.

We need to be careful about the domain of the operator on the right-hand side of (\ref{3eq5.35a}). 
First, since ${\cal L}$ is an unbounded operator, it is necessary to check that the right-hand side of (\ref{3eq6.11}) is finite for each $n$. It can be shown \cite{generalized}
 that all the coefficients $\bar r _n$ given by (\ref{1eq6.18}) are finite if
\numparts
\begin{equation}
\label{1eq6.23}
V(-\infty)=\pm \infty, \qquad 
\lim_{x \to -\infty}\frac{\log \vert x \vert}{V(x)}=0, 
\end{equation}
or
\begin{equation}
\label{1eq6.24}
V(-\infty)=V_0, \qquad 
\lim_{x \to -\infty} \vert x \vert^n [V(x)-V_0]=0 \quad {\rm for\  any} \ n.
\end{equation}
\endnumparts
Second, for the right-hand side of (\ref{3eq6.8}) to make sense, the function ${\cal A}\,\bar r_{N+1}(x,W)$ must lie in the domain of $({\cal A}-2ik{\cal B})^{-1}$.
As shown in appendix~E, this requirement, too, is satisfied if either (\ref{1eq6.23}) or (\ref{1eq6.24}) holds. 
Aside from these two points, there is no problem in (\ref{3eq6.6}). Expression (\ref{3eq6.6}) is correct for any nonnegative integer $N$ as long as the potential satisfies either (\ref{1eq6.23}) or (\ref{1eq6.24}).

If the remainder term satisfies
\begin{equation}
\label{1eq6.30}
\lim_{k \to 0}\frac{ {\bar \rho}_N}{k^N}=0
\end{equation}
for any $N$, then (\ref{3eq6.6}) gives the asymptotic expansion
\begin{equation}
\label{1eq6.31}
\bar R_r(x,-\infty)=\bar r_0+ ik \bar r_1 + (ik)^2 \bar r_2 + (ik)^3 \bar r_3 +\cdots.
\end{equation}
(In this paper we use the term ^^ ^^ asymptotic" in a broader sense, including the convergent cases.) 
The expansion of the original $R_r$ is obtained from (\ref{1eq6.31}) 
by setting $W=V(x)$:
\begin{equation}
\label{1eq6.33}
R_r(x,-\infty)=r_0+ ik r_1 + (ik)^2 r_2 + (ik)^3 r_3 +\cdots,
\end{equation}
where $
r_n(x)\equiv \bar r_n(x,W=V(x)).
$
The explicit forms of the first few coefficients are
\numparts
\begin{eqnarray}
\fl
\label{1eq6.20}
r_1=2e^{V(x)} [\hbox{$-$}]_{-\infty}^x,
\qquad 
r_2=4 e^{2V(x)} [\hbox{$-$}\hbox{$-$}]_{-\infty}^x, 
\nonumber \\
\fl
r_3 =12 e^{3V(x)} [\hbox{$-$}\hbox{$-$}\hbox{$-$}]_{-\infty}^x
-4e^{V(x)} [\hbox{$-$}\hbox{$-$}\hbox{$+$}]_{-\infty}^x,
\qquad {\rm for} \quad V(-\infty)=+\infty, 
\end{eqnarray}
\begin{eqnarray}
\fl
\label{1eq6.21}
r_1=-2e^{-V(x)} [\hbox{$+$}]_{-\infty}^x,
\qquad
r_2=-4 e^{-2V(x)} [\hbox{$+$}\hbox{$+$}]_{-\infty}^x, 
\nonumber \\
\fl
r_3=-12 e^{-3V(x)} [\hbox{$+$}\hbox{$+$}\hbox{$+$}]_{-\infty}^x
+4e^{-V(x)} [\hbox{$+$}\hbox{$+$}\hbox{$-$}]_{-\infty}^x,
\qquad {\rm for} \quad  V(-\infty)=-\infty, 
\end{eqnarray}
\begin{eqnarray}
\fl
\label{1eq6.22}
r_1=\case{1}{2}\, {\rm sech}^2\,\case{V(x)-V_0}{2} \,(\pm\,]_{-\infty}^x, 
\nonumber \\
\fl
r_2=\case{1}{2}{\rm sech}^3\,\case{V(x)-V_0}{2}
 \Bigl\{
 e^{[V_0+V(x)]/2}(\hbox{$\pm$}\hbox{$-$}]_{-\infty}^x
 +e^{-[V_0+V(x)]/2} &(\hbox{$\pm$}\hbox{$+$}]_{-\infty}^x 
\Bigr\}, \qquad {\rm for} \quad  V(-\infty)=V_0.
\nonumber \\
\end{eqnarray}
\endnumparts
Here we have written, for simplicity,  $[\hbox{$-$$-$$+$}]_{-\infty}^x$ etc 
in place of $[-1,-1,+1]_{-\infty}^x$ etc.

It remains for us to study whether (\ref{1eq6.30}) holds or not.
If we assume that
\begin{equation}
\label{2eq5.30}
\lim_{k \to 0} \frac{1}{{\cal A}-2ik{\cal B}}\,g={\cal A}^{-1}g,
\end{equation}
 then from (\ref{3eq6.8}) it follows that 
 \begin{equation}
 \label{1eq6.39}
\lim_{k \to 0}\frac{1}{(ik)^{N+1}} \bar \rho_N
=\lim_{k \to 0}\frac{1}{{\cal A}-2ik{\cal B}}
\,{\cal A}\,\bar r_{N+1}
={\cal A}^{-1}{\cal A}\,\bar r_{N+1} =\bar r_{N+1},
\end{equation}
and so (\ref{1eq6.30}) holds as long as $\bar r_{N+1}$ is finite. 
However, since the limit and the integral are not necessarily interchangeable, there is no guarantee for (\ref{2eq5.30}).
We need to check whether the second equality of (\ref{1eq6.39}) really holds.
This can be done by using expressions (\ref{1eq6.25}) for the remainder term given in appendix~D. It is shown in appendix~F that (\ref{1eq6.39}) is indeed true as long as $\bar r_{N+1}$ is finite. Therefore, the asymptotic expansion (\ref{1eq6.31}) is valid if $\bar r_n$ is finite for any $n$, i.e., if (\ref{1eq6.23}) or (\ref{1eq6.24}) is satisfied.
In other words, the reflection coefficient $R_r$ can be asymptotically expanded in the form of (\ref{1eq6.33}) if  $V(x)$ tends to infinity more rapidly than logarithmically or 
converges to $V_0$ more rapidly than any power of $x$ as $x \to -\infty$. 

Finally, let us comment on the convergence property of the series (\ref{1eq6.33}). 
Here we omit the explanation, but it can be shown that the power series (\ref{1eq6.33}) has a nonzero radius of convergence if 
$f(-\infty) \neq 0$, i.e., if $V(x)$ diverges linearly or faster as $x \to -\infty$. 
If $V(-\infty)=V_0$ is finite, (\ref{1eq6.33}) is convergent for small  $\vert k \vert$ provided that $V(x)$ tends to $V_0$ exponentially or faster as $x \to -\infty$. 
 (See example~4 of section~7.) 

If $V(x)$ diverges more slowly than $\vert x \vert$ and more rapidly than $\log \vert x \vert$, 
or if $V(x)$ converges to $V_0$ slower than exponentially and faster than any power of $\vert x \vert$, 
then the series (\ref{1eq6.33}) is asymptotic but divergent. In such cases $R_r(k)$ is essentially singular at $k=0$.  (See example~6 of section~7). 

If $V(x)$ diverges logarithmically or more slowly,  
or if $V(x)$ tends to $V_0$ with a power law or more slowly, then the small-$k$ behavior of $R_r$ cannot be expressed as an asymptotic series of the form of (\ref{1eq6.33}).  (See example~7 of section~7. The Schr\"odinger potentials studied by Klaus in \cite{klaus} correspond to the marginal case.)


\section{High-energy expansion}
Substituting (\ref{3eq5.35b}) into (\ref{1eq5.19}), we obtain
\begin{equation}
\label{1eq7.3}
\bar R_r(x,-\infty)=\bar  c_0 + \frac{1}{2ik}\bar c_1+ \frac{1}{(2ik)^2} \bar c_2 + \cdots+\frac{1}{(2ik)^N}\bar c_N+\bar \delta_N,
\end{equation}
where
\refstepcounter{equation}
\label{1eq7.4}
\addtocounter{equation}{-1}
\begin{eqnarray}
\bar c_0\equiv -\xi, 
\qquad
\bar c_n\equiv -(2 {\cal L}^{-1})^{n-1}(1-\xi^2)f(x)
\quad (n \geq 1),
\end{eqnarray}
\begin{equation}
\label{1eq7.5}
\bar \delta_N=\frac{-1}{(2ik)^N}\frac{1}{{\cal A}-2ik{\cal B}}\,{\cal B}\,\bar c_{N+1}.
\end{equation}
(Here we used ${\cal B}^{-1}(1-\xi^2)f(x)=(1-\xi^2)f(x)$.)
From (\ref{1eq5.1}) and (\ref{1eq5.25}) we have
\begin{equation}
\label{1eq7.6}
{\cal L}^{-1}g(x,\xi)=\frac{1-\xi^2}{2\xi}\int_0^\xi\left[\frac{1}{1-\xi^2}\frac{\partial}{\partial x}
+f(x)\frac{\partial}{\partial \xi}\right]g(x,\xi)\,d\xi.
\end{equation}
To calculate the $\bar c_n$, it is convenient to define
\begin{equation}
\label{1eq7.9}
{\tilde c}_n(x,\xi) \equiv \frac{{\bar c}_n(x,\xi)}{1-\xi^2}
\qquad (n \geq 1),
\end{equation}
and rewrite the second equation of (\ref{1eq7.4}) in the form
\begin{equation}
\label{2eq7.8}
\tilde c_n=-{\cal M}^{n-1} f,
\qquad
{\cal M} \equiv \frac{2}{1-\xi^2}\,{\cal L}^{-1} (1-\xi^2).
\end{equation}
The operator ${\cal M}$ acts as
\begin{equation}
\label{2eq7.10}
\fl
{\cal M}g(x,\xi)=f(x)\left[
\frac{g(x,\xi)-g(x,0)}{\xi}-\xi g(x,\xi)\right]
+\frac{1}{\xi}\int_0^\xi \frac{\partial }{\partial x}g(x,\xi)\,d\xi.
\end{equation}
Using (\ref{2eq7.10}) successively in the first equation of (\ref{2eq7.8}), we obtain
\begin{eqnarray}
\label{1eq7.10}
{\tilde c}_1=-f, \qquad
{\tilde c}_2=-f'+f^2 \xi, \qquad
{\tilde c}_3=-f''+f^3+2ff'\xi - f^3 \xi^2,
\nonumber\\
{\tilde c}_4=-f'''+5f^2f'-(2f^4-{f'}^2-2ff'')\xi -3f^2f'\xi^2 +f^4\xi^3,
 \quad {\rm etc.}
\end{eqnarray}
The $\tilde c_n$ are $(n-1)$ th order polynomials in $\xi$. 
The $\bar c_n$ are obtained as $\bar c_n =(1-\xi^2) \tilde c_n$. 

Using (\ref{1eq5.13}), expression (\ref{1eq7.5}) for the remainder term can be rewritten as
\begin{equation}
\label{1eq7.12}
\bar \delta_N
=\frac{1}{(2ik)^N}\int_{-\infty}^x
\bar \tau^2(x,z)K_N\left(z,\bar R_l(x,z)\right)\,dz,
\end{equation}
\begin{equation}
\label{1eq7.13}
K_n(x,\xi) \equiv -\left(1+\xi \frac{\partial}{\partial \xi}\right)\tilde c_{n+1}(x,\xi).
\end{equation}

Expression (\ref{1eq7.3}) makes sense if and only if the $\bar c_n$ and the $\bar \delta_n$ given by (\ref{1eq7.4}) and (\ref{1eq7.5}) are finite.
We can easily see that $\bar c_n$ contains derivatives of $f$ up to $f^{(n-1)}$.
 So $\bar c_n$ is finite if $f(x)$ is ($n-$1)-times differentiable.
We can also show that (\ref{1eq7.5}) makes sense and is finite if $f^{(N-1)}(x)$ is continuous and piecewise differentiable. (See appendix~E.)
Therefore, expression (\ref{1eq7.3}) is correct provided that $f(x)$ is ($N-$1)-times continuously differentiable and that $f^{(N-1)}(x)$ is piecewise differentiable.

The expansion of the original $R_r$ is obtained from (\ref{1eq7.3}) 
by setting $\xi=0$:
\begin{equation}
\label{1eq7.15}
\fl
R_r(x,-\infty;k)=\frac{1}{2ik}c_1(x)+ \frac{1}{(2ik)^2}c_2(x) + 
\cdots+\frac{1}{(2ik)^N}c_N(x)+\delta_N(x,k),
\end{equation}
where $
c_n(x)\equiv \bar c_n(x,\xi=0) =\tilde c_n(x,\xi=0).
$
From (\ref{1eq7.10}) we find
\begin{equation}
\label{1eq7.17}
\fl
c_1=-f, \qquad c_2=-f', \qquad c_3=-f''+f^3, \qquad
c_4=-f'''+5f^2f', \qquad {\rm etc.}
\end{equation}
The $\delta_N$ in (\ref{1eq7.15}) is obtained from (\ref{1eq7.12}) by replacing $\bar \tau$ 
and $\bar R_l$ with $\tau$ and $R_l$:
\begin{equation}
\label{1eq7.18}
\delta_N(x,k)
=\frac{1}{(2ik)^N}\int_{-\infty}^x
\tau^2(x,z;k)K_N\left(z,R_l(x,z;k)\right)\,dz.
\end{equation}
If this remainder term satisfies
\begin{equation}
\label{1eq7.19}
\lim_{\vert k\vert \to \infty} k^N \delta_N(x,k)=0,
\end{equation}
then (\ref{1eq7.15}) can be written as
\begin{equation}
\label{2eq8.1}
\fl
R_r(x,-\infty;k)=\frac{1}{2ik}c_1(x)+ \frac{1}{(2ik)^2}c_2(x) + 
\cdots+\frac{1}{(2ik)^N}c_N(x)+o(1/\vert k \vert^N).
\end{equation}
If (\ref{1eq7.19}) holds for any $N$, then we have the asymptotic expansion
\begin{equation}
\label{1eq7.20}
R_r(x,-\infty;k)
=\frac{1}{2ik}c_1+ \frac{1}{(2ik)^2} c_2 + \frac{1}{(2ik)^3} c_3 +\frac{1}{(2ik)^4} c_4 +\cdots.
\end{equation} 
In the next section we shall study the conditions for (\ref{1eq7.19}) to hold.
(Note that (\ref{1eq7.19}) is equivalent to 
$\lim_{\vert k \vert \to \infty} k^N \bar \delta_N(x,k)=0$, as is obvious from the definition of $\bar R_r$.)


\section{Validity of (\ref{1eq7.19})}

\noindent
The condition for the validity of (\ref{1eq7.19}) differs depending on 
 the way how we let $\vert k \vert$ go to infinity. 
 We consider the following three ways of taking this limit:
 
\medskip
 (i)\quad $\vert k \vert \to \infty$ with fixed $\arg k$,\, $0<\arg k<\pi$,
 
\smallskip
 (ii) \quad $\vert k \vert\to \infty$ with fixed ${\rm Im}\,k >0$,
 
\smallskip
 (iii) \quad $\vert k \vert \to \infty$ with ${\rm Im}\,k=0$  (i.e., $\arg k=0$ or $\pi$).
 
\medskip
\noindent 
In this section we shall show that:
\begin{itemize}
\item In case (i), equation (\ref{1eq7.19}) holds for any $f(x)$ as long as $f^{(N-1)}(x)$ is continuous and piecewise differentiable. 
\item
In case (ii), equation (\ref{1eq7.19}) holds if $f^{(N-1)}(x)$ is continuous and piecewise differentiable, and if $f(x)$ does not diverge exponentially or faster as $x \to -\infty$.
\item
In case (iii), equation (\ref{1eq7.19}) holds if $f^{(N-1)}(x)$ is continuous and piecewise differentiable, and if $f(-\infty)$ is finite.
\end{itemize}
(It is always assumed that $f(x)$ satisfies the conditions stated in the introduction.
Recall that (\ref{1eq7.18}) is well defined if $f^{(N-1)}$ is continuous and piecewise differentiable. )

Let us assume that $f(x)$ is continuous. 
When both $x$ and $y$ are finite, we can easily show, for ${\rm Im}\,k \geq 0$, 
that $\tau(x,y;k)$ and $R_l(x,y;k)$ have the following properties:
\begin{equation}
\fl
\label{b1eqg.1}
\tau(x,y;k) = e^{ik(x-y)}[1+O(1/\vert k \vert)],
\qquad
R_l(x,y;k)=O(1/\vert k \vert)
\qquad 
{\rm as} \quad  \vert k \vert \to \infty,
\end{equation}
\begin{equation}
\label{b1eqg.2}
\vert \tau(x,y;k) \vert \leq e^{-{\rm Im}\,k(x-y)},
\qquad
\vert R_l(x,y;k)\vert \leq 1.
\end{equation}
(See, for example, \cite{theory} and references therein.)
We shall use (\ref{b1eqg.1}) and (\ref{b1eqg.2}) in our proof. 

Since $K_n(x,\xi)$ is an $n$th order polynomial in $\xi$, 
we may write
\begin{equation}
\label{1eq8.1}
K_n(x,\xi)\equiv \sum_{m=0}^n \xi^m h_{nm}(x),
\end{equation}
where $h_{nm}$ are polynomials in $f$ and its derivatives.
Their explicit forms are
\begin{eqnarray}
\label{1eq8.2}
\fl
h_{00}=f, \quad h_{10}=f', \quad h_{11}=-2f^2, \quad
h_{20}=f''-f^3, \quad h_{21}=-4ff', \quad h_{22}=3f^3,
\nonumber \\
\fl
h_{30}=f'''-5f^2f', \quad h_{31}=4f^4-2f'^2-4ff'', \quad 
h_{32}=9f^2f', \quad h_{33}=-4f^4.
\end{eqnarray}
Obviously (\ref{1eq7.19}) is satisfied if, for any $m \leq N$,
\begin{equation}
\label{1eq8.4}
\lim_{\vert k \vert \to \infty}
\int_{-\infty}^x \tau^2(x,z;k) R_l^m(x,z:k)h_{Nm}(z)\,dz=0.
\end{equation}
Now let us show that (\ref{1eq8.4}) holds for any $m \leq N$ under the conditions listed above.

\bigskip
\noindent
(i) \, $\vert k \vert \to \infty$ with fixed $\arg k$ $( 0<\arg k <\pi)$. 

\nobreak
\medskip
\noindent
In this case, both $\tau(x,z;k)$ and $R_l(x,z;k)$ vanish as $\vert k \vert \to \infty$, as can be seen from (\ref{b1eqg.1}).
So, if it is possible to interchange the limit and the integral as
\begin{equation}
\label{1eq8.6}
\lim_{\vert k \vert \to \infty}
\int_{-\infty}^x\tau^2 R_l^m h_{Nm}\,dz=
\int_{-\infty}^x \lim_{\vert k \vert \to \infty} \tau^2 R_l^m h_{Nm}\,dz
\end{equation}
then (\ref{1eq8.4}) holds, since the right-hand side of (\ref{1eq8.6}) is obviously zero. 
Since $\vert R_l \vert \leq 1$, 
equation (\ref{1eq8.6}) holds if there exist a $k$-independent real function $A(z)$ 
and a real number $a$ such that $
\vert \tau^2(x,z;k)\vert \leq A(z)$ for $\vert k \vert \geq a$, 
and $\int_{-\infty}^xA(z)\vert h_{Nm}(z)\vert\,dz<\infty$. 
It is always possible to find such $A(z)$ and $a$.
(If $f(z)$ diverges as $z \to -\infty$ exponentially or more rapidly, we have    $A(z)=C\exp[-\vert V(z)\vert]$ with a constant $C$. Otherwise, we may take $A(z)=e^{-2a\sin \theta(x-z)}$, $\theta=\arg k$.) 
So (\ref{1eq8.4}) holds for any $f(x)$ as long as $h_{Nm}$ is finite.

\bigskip
\noindent
(ii) \, $\vert k \vert \to \infty$ with fixed ${\rm Im}\,k>0$.

\nobreak
\medskip
\noindent
Let $b \equiv {\rm Im}\,k>0$. 
In this case, $\tau(x,z;k)$ approaches $e^{ik(x-z)}$ as $\vert k \vert \to \infty$. 
(See (\ref{b1eqg.1}).) 
Let us first consider the case $m=0$ in (\ref{1eq8.4}).
From (\ref{b1eqg.2}) it is obvious that 
$
\left\vert
\tau^2(x,z;k)-e^{2ik(x-z)}
\right\vert \leq2 e^{-2b(x-z)}.
$ 
So, if 
$
\int_{-\infty}^x \vert h_{N0}(z)\vert e^{2bz} \,dz<\infty,
$
then it is permissible to replace the $\tau^2$ by $e^{2ik(x-z)}$ within the integral: 
\begin{equation}
\label{1eq8.18}
\fl
\lim_{\vert k \vert \to  \infty}
\int_{-\infty}^x \tau^2(x,z)h_{N0}(z)\,dz
=\lim_{\vert {\rm Re\,} k \vert \to  \infty}
\int_{-\infty}^x 
e^{-2i({\rm Re\,}k)(z-x)}
e^{2b(z-x)}
h_{N0}(z)\,dz.
\end{equation}
The right-hand side of (\ref{1eq8.18}) vanishes according to the Riemann-Lebesgue theorem.  

In the same way, we can show that (\ref{1eq8.4}) holds for any $m$ if
$
\int_{-\infty}^x \vert h_{Nm}(z)\vert e^{2bz} \,dz<\infty.
$
(The problem is easier for $m\neq 0$, since $R_l^m \to 0$ as $\vert k \vert \to \infty$.) 
Since $h_{Nm}$ is a polynomial in $f$ and its derivatives, this condition is satisfied for any $\{n,m\}$ if $f(-\infty)$ is finite, or if $f(z)$ tends to infinity more slowly than any exponential function as $z \to -\infty$. 

\bigskip
\noindent
(iii) \, $\vert k \vert \to \infty$ with ${\rm Im}\, k=0$.

\nobreak
\medskip
\noindent
The above argument is also applicable to the case $b=0$. If
$
\int_{-\infty}^x \vert h_{Nm}(z)\vert \,dz<\infty,
$
then (\ref{1eq8.4}) holds. 
If $f(z)$ falls off with a power law or faster, this condition is satisfied for sufficiently large $N$. 
This implies that (\ref{1eq7.19}) holds for any $N$ for such $f(z)$. 

If $f(z)$ goes to zero more slowly than any power of $\vert z\vert$, 
then (\ref{1eq8.4}) cannot be proved by the above method. 
However, (\ref{1eq8.4}) holds in this case, too. 
When $(-z)$ is large, $\tau^2(x,z) R_l^m(x,z)$ has the approximate form (see (\ref{1eq3.8}) and (\ref{1eq3.11}) with $\xi=0$)
\begin{equation}
\label{1eq8.23}
\tau^2(x,z) R_l^m(x,z)
\simeq 
C^2(x,k) D^m(x,k) e^{2(1+m)i[-kz +\theta(z,k)]},
\end{equation}
where $\theta(z,k)$ is a real function which is $o(\vert z\vert)$ as $z \to -\infty$ 
and $o(1)$ as $\vert k \vert \to \infty$.
It can shown that $\vert C(x,k)\vert=1+O(1/\vert k \vert^2)$ and $D(x,k)=O(1/\vert k\vert)$ 
as $\vert k \vert \to \infty$.
For sufficiently large $(-z_1)$, we can evaluate
\begin{equation}
\label{1eq8.24}
\fl
\int_{-\infty}^{z_1}\tau^2 R_l^m h_{nm}\,dz
\simeq
\frac{iC^2(x,k)D^m(x,k) }{2(1+m)k}e^{2(1+m)i[-kz_1+\theta(z_1,k)]}h_{nm}(z_1).
\end{equation}
The right-hand side vanishes like $1/\vert k \vert^{1+m}$ in the limit $\vert k \vert \to \infty$.  
Whereas (\ref{1eq8.23}) is an approximation, we can show that
the part omitted on the right-hand side of (\ref{1eq8.24}) are of higher order than $1/\vert k \vert^{1+m}$. Hence we may conclude that (\ref{1eq8.4}) holds if $f(-\infty)=0$.

In the same way, it can be shown that (\ref{1eq8.4}) also holds when  $f(-\infty)=c \,
(\neq 0, \pm\infty)$. 
(In this case $R_r$ has branch cuts along the real axis. So we need to replace $k$ by $k+i\epsilon$ with positive $\epsilon$, and let $\epsilon \to 0$ after evaluating the integral.)

\bigskip
Thus, we have shown that (\ref{1eq7.19}) holds under the stated conditions.
The conditions for the validity of the asymptotic expansion (\ref{1eq7.20}) are obtained by replacing the phrase ^^ ^^ $f^{(N-1)}(x)$ is continuous and piecewise differentiable"  by ^^ ^^ $f(x)$ is infinitely differentiable".  
Let us remark that these are sufficient conditions, not necessary ones. 
There are cases where (\ref{1eq7.20}) is valid even though $f(x)$ is not infinitely differentiable, and even though (\ref{1eq7.18}) is not well defined. 
When the potential is a piecewise analytic function, the expansion (\ref{1eq7.20}) is correct for $0<\arg k<\pi$ if the point $x$ is away from the singularities. In such cases the effect of the singularities falls off exponentially as ${\rm Im}\,k\to \infty$, and so (\ref{1eq7.20}) is not affected. 
(See example~8 of section~7.)


\section{Examples}
For some simple potentials, it is possible to obtain the exact form of $R_r(x,-\infty;k)$. 
In this section, we shall compare the exact expressions with the results of 
our high-energy and low-energy expansions.
We omit the derivation of the exact results.

\medskip
\noindent
{\bf Example 1}\,: \quad $V(x)=-2 x, \quad f(x)=1.$

\nobreak
\medskip
\noindent
Our first example is a linear potential. The exact form of $R_r$ for this $V(x)$ is
\begin{equation}
\label{1eq9.1}
R_r(x,-\infty;k)=ik + \sqrt{1-k^2} = ik\left[1-\sqrt{1-(1/k)^2}\right],
\end{equation}
which is independent of $x$. 
Since $f(x)$ is a constant, the $c_n$ obtained from (\ref{1eq7.17}) are also $x$-independent. Obviously $c_2=c_4=c_6=\cdots=0$, and (\ref{1eq7.20}) reads
\begin{equation}
\label{1eq9.2}
R_r= -\frac{1}{2ik}+ \frac{1}{(2ik)^3}-\frac{2}{(2ik)^5}+\frac{5}{(2ik)^7}
-\frac{14}{(2ik)^9}+ \cdots.
\end{equation}
It is obvious that (\ref{1eq9.2})  is the correct expansion of (\ref{1eq9.1}). 
Since the only singularities of (\ref{1eq9.1}) are 
the branch points at $k=\pm 1$, the series (\ref{1eq9.2}) is convergent for $\vert k \vert >1$.

Putting $V(x)=-2x$ in definition (\ref{1eq6.1}), we have
$ [\hbox{$-$}]_{-\infty}^x=\frac{1}{2}e^{2x}$, 
$ [\hbox{$-$}\hbox{$-$}]_{-\infty}^x 
=\frac{1}{2}([\hbox{$-$}]_{-\infty}^x)^2=\frac{1}{8} e^{4x}$, 
$ [\hbox{$-$}\hbox{$-$}\hbox{$+$}]_{-\infty}^x =\frac{1}{16} e^{2x}$, 
and so on. Substituting them in (\ref{1eq6.20}), we obtain the low-energy expansion
\begin{equation}
\label{1eq9.3}
R_r= 1+ ik + \frac{1}{2} (ik)^2-\frac{1}{8}(ik)^4+\frac{1}{16}(ik)^6+\cdots,
\end{equation}
which is obviously the correct expansion of (\ref{1eq9.1}). This series is convergent 
for $\vert k \vert<1$.  

\medskip
\noindent
{\bf Example 2}\,: \quad $V(x)=x^2, \quad f(x)=-x.$

\nobreak
\medskip
\noindent
The next example is a parabolic potential.
The exact form of the reflection coefficient for this potential can be expressed in terms of the confluent hypergeometric function 
$F(\alpha,\gamma;z)
=\sum_{n=0}^\infty
\frac{\alpha(\alpha+1)\cdots(\alpha+n-1)}{\gamma(\gamma+1)\cdots(\gamma+n-1)}
\frac{1}{n!}z^n
$ and the gamma function. We have
\begin{eqnarray}
\label{1eq9.5}
R_r(x,-\infty;k)=\frac{a(x,k)}{a(x,-k)},
\\
\fl
a(x,k)\equiv 
\Gamma\left(1-\case{k^2}{4}\right) 
\left[
F\left(-\case{k^2}{4}, \case{1}{2};x^2\right)
 + ikx F\left(1-\case{k^2}{4}, \case{3}{2};x^2\right) 
\right]
\nonumber \\
\fl
\qquad \qquad \quad
+ (ik/2)
\Gamma\left(\case{1}{2}-\case{k^2}{4}\right) 
\left[
F\left(\case{1}{2}-\case{k^2}{4}, \case{1}{2};x^2\right) 
+ ikx F\left(\case{1}{2}-\case{k^2}{4}, \case{3}{2};x^2\right) 
\right].
\end{eqnarray}
The high-energy expansion obtained from (\ref{1eq7.20}) and (\ref{1eq7.17}) is
\begin{equation}
\label{1eq9.6}
R_r = \frac{x}{2ik}+\frac{1}{(2ik)^2}-\frac{x^3}{(2ik)^3}-\frac{5x^2}{(2ik)^4}
+\frac{2x^5-11x}{(2ik)^5}+\cdots.
\end{equation}
Using the asymptotic forms of $F(\alpha,\gamma;z)$ and $\Gamma(z)$,  
we can show that (\ref{1eq9.6}) is the correct asymptotic form of (\ref{1eq9.5}) as 
$\vert k \vert \to \infty$ with $0<\arg k <\pi$ (figure~1(a)).
\begin{figure}
\hspace{1cm}
\includegraphics[scale=0.7]{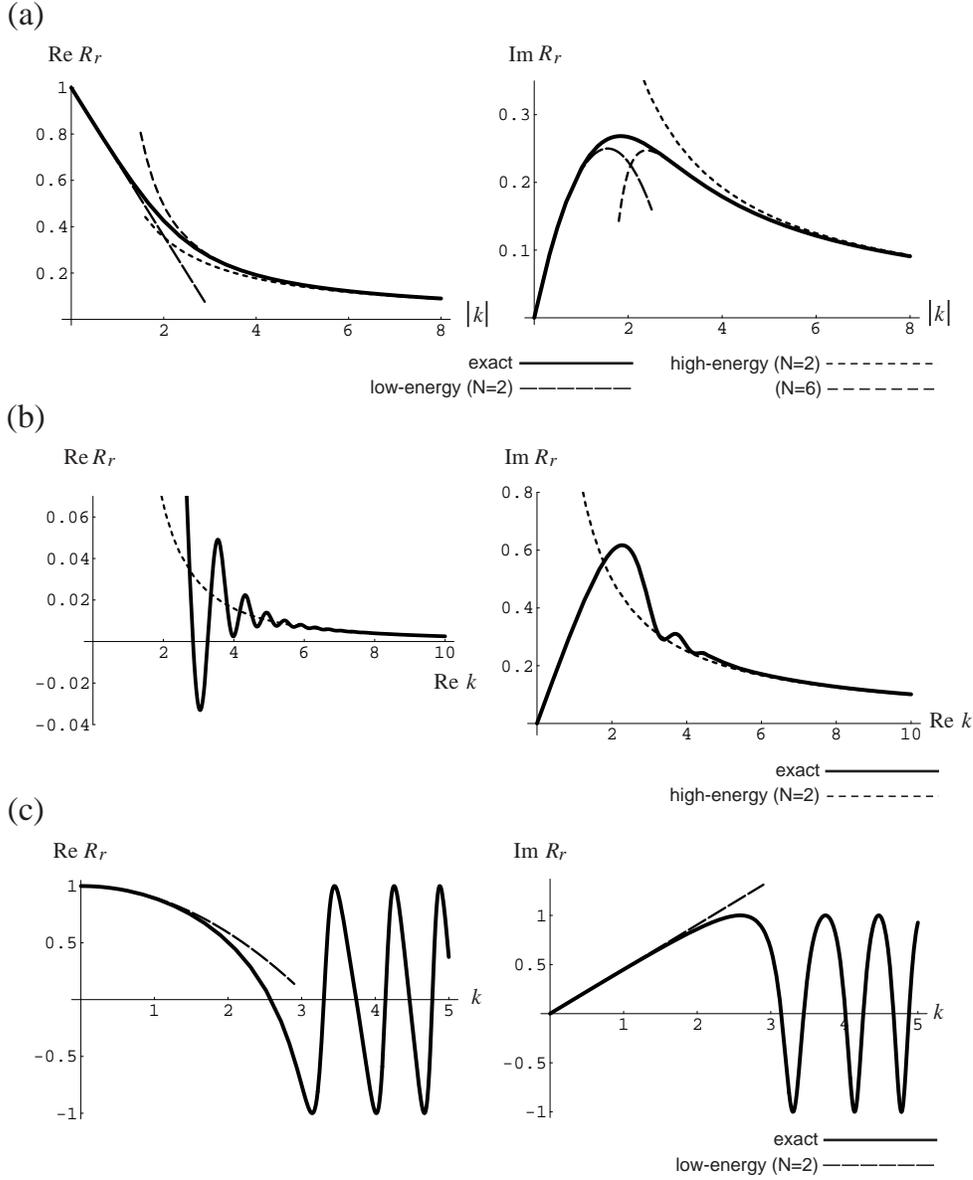}
\caption{
The real and imaginary parts of $R_r(x_0,-\infty;k)$ for the potential $V(x)=x^2$ (example~2), 
where $x_0=-2$.
They are plotted along three different lines in the complex $k$ plane: 
(a) $\arg k=\pi/4$; \,
(b) ${\rm Im}\,k=1/2$; \,
(c) ${\rm Im}\,k=0$.  \,
In (a) and (b), the abscissa is $\vert k \vert$ and ${\rm Re }\,k$, respectively. 
Solid lines: the exact $R_r$ (equation (\ref{1eq9.5})); \
broken lines: the low-energy expansion (\ref{1eq9.7}) up to order $k^2$;\ 
dashed lines: the high-energy expansion (\ref{1eq9.6}) up to order $1/k^N$ ($N=2$ and $6$).
}
\end{figure}
The series (\ref{1eq9.6}) is divergent in this case.
This asymptotic expression also holds in the case where $\vert k \vert \to \infty$ 
with fixed ${\rm Im}\,k>0$ (figure~1(b)).  However, (\ref{1eq9.6}) does not hold when ${\rm Im}\,k=0$. In that case the exact $R_r(k)$ oscillates and does not tend to zero as $\vert k \vert \to \infty$ (figure~1(c)).

The coefficients of the low-energy expansion of $R_r$ for this potential are also obtained from (\ref{1eq6.20}). We have 
$ [\hbox{$-$}]_{-\infty}^x=\int_{-\infty}^x e^{-z^2}dz=(\sqrt{\pi}/2){\rm erfc\,}(-x)$, 
where ${\rm erfc\,}z$ is the Gauss error function.
Substituting this into (\ref{1eq6.20}), we obtain
\begin{equation}
\label{1eq9.7}
R_r= 1+\sqrt{\pi}e^{x^2}{\rm erfc}\,(-x) ik
+ \frac{\pi}{2}e^{2x^2}[{\rm erfc}\,(-x)]^2 (ik)^2+ \cdots.
\end{equation}
(See figure~1.) 
The $R_r(k)$ given by (\ref{1eq9.5}) has poles in the lower half plane. 
The series (\ref{1eq9.7}) is convergent if $\vert k \vert$ is smaller than the 
distance from the origin to the nearest pole.

\medskip
\noindent
{\bf Example 3}\,: \quad $V(x)=e^{-x}, \quad f(x)=\frac{1}{2}e^{-x}.$

\nobreak
\medskip
\noindent
This is an exponential potential, which tends to infinity as $x \to -\infty$ more rapidly than the previous examples. The exact $R_r$ has the form
\begin{equation}
\label{1eq9.8}
\fl
R_r(x,-\infty;k)=
i \frac{J_{-\nu}(-ie^{-x}/2)-ie^{-k\pi}J_{\nu}(-ie^{-x}/2)}
{J_{1-\nu}(-ie^{-x}/2)+ie^{-k\pi}J_{\nu-1}(-ie^{-x}/2)},
\qquad \nu \equiv ik+\frac{1}{2},
\end{equation}
where $J_\alpha(z)$ is the Bessel function.
The high-energy expression (\ref{1eq7.20}) now reads 
\begin{equation}
\label{1eq9.9}
R_r=-\frac{e^{-x}}{4ik}+\frac{e^{-x}}{2(2ik)^2}
-\frac{4 e^{-x}-e^{-3x}}{8(2ik)^3}
+\frac{4 e^{-x} -5 e^{-3x}}{8(2ik)^4}+\cdots,
\end{equation}
As $\vert k \vert \to \infty$ with $\arg k$ fixed in the region $0<\arg k <\pi$, 
the $R_r$ given by (\ref{1eq9.8}) 
has the asymptotic form (\ref{1eq9.9}). However, this expression does not hold
when ${\rm Im}\,k$ is kept fixed. 
 
Using (\ref{1eq6.20}) we obtain the low-energy expansion for this potential as
\begin{equation}
\label{1eq9.10}
\fl
R_r= 1-2\exp(e^{-x}){\rm Ei}\,(-e^{-x})ik
+2 \exp(2e^{-x})[{\rm Ei}\,(-e^{-x})]^2(ik)^2 + \cdots,
\end{equation}
where ${\rm Ei}\,(z)=-\int_{-z}^\infty (1/t)e^{-t}dt$ is the exponential integral function.

\medskip
\noindent
{\bf Example 4}\,: \quad $V(x)=e^x, \quad f(x)=-\frac{1}{2}e^x.$

\nobreak
\medskip
\noindent
This potential falls off rapidly as $x \to -\infty$. The exact $R_r$ for this $V(x)$ is 
\begin{equation}
\label{1eq9.11}
R_r(x,-\infty;k)=-i \frac{J_{1-\nu} (-ie^x/2)}{J_{-\nu}(-ie^x/2)}, \qquad \nu \equiv ik+\frac{1}{2}.
\end{equation} 
From (\ref{1eq7.20}) and (\ref{1eq7.17}) we have
\begin{equation}
\label{1eq9.12}
R_r=\frac{e^{x}}{4ik}+\frac{e^{x}}{2(2ik)^2}
+\frac{4 e^{x}-e^{3x}}{8(2ik)^3}
+\frac{4 e^{x} -5 e^{3x}}{8(2ik)^4}+\cdots.
\end{equation}
This is the correct asymptotic expansion of (\ref{1eq9.11}). 
Unlike the previous example, this expansion is valid even when ${\rm Im}\,k=0$. (See figure~2(a).)

\begin{figure}
\hspace{1cm}
\includegraphics[scale=0.7]{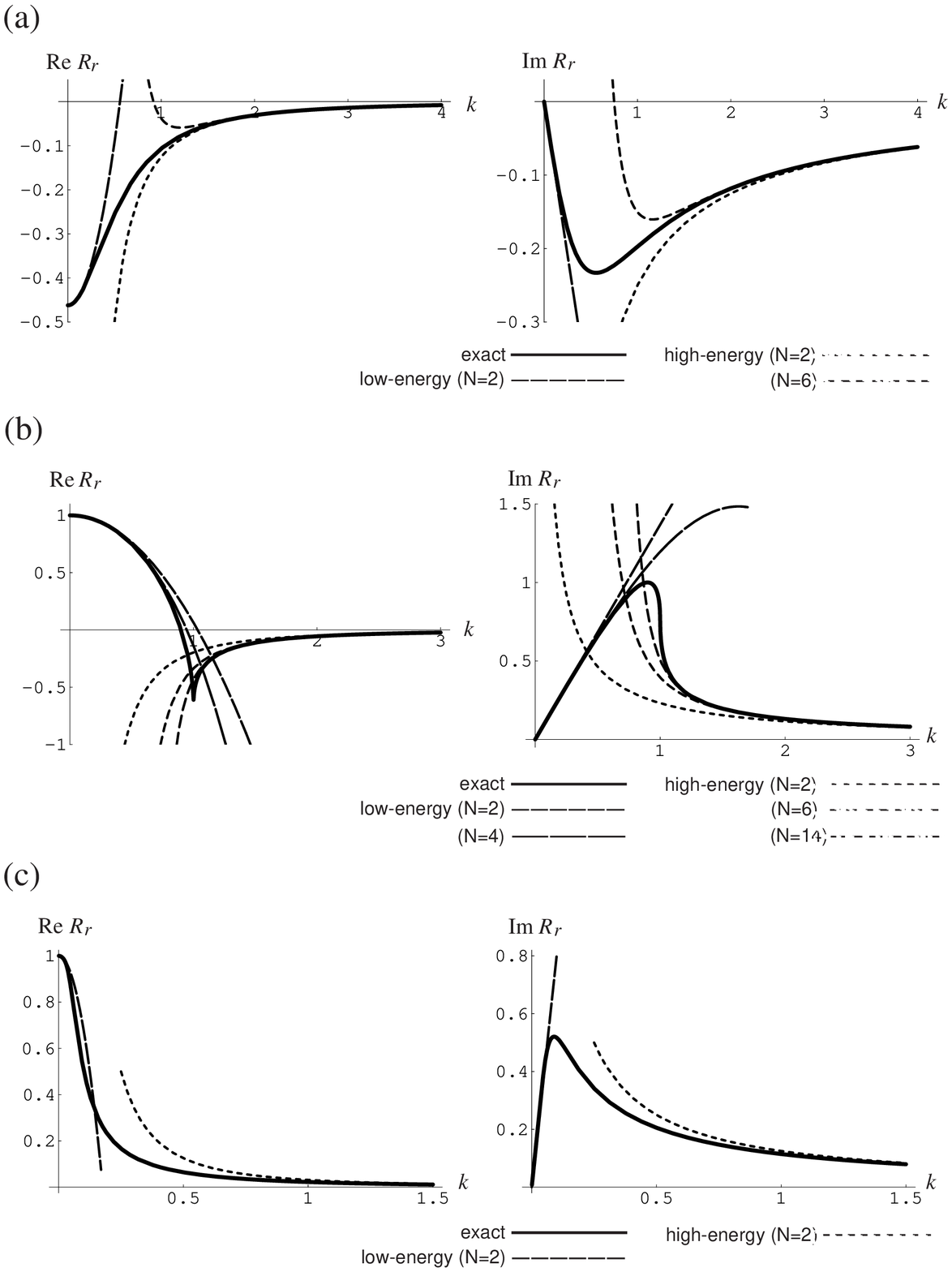}
\caption{
The real and imaginary parts of $R_r(x_0,-\infty;k)$ as functions of real $k$.

\noindent
(a) \ (example~4)\  $V(x)=e^{-x}$; $x_0=0$. 
Solid lines: the exact $R_r$ (equation (\ref{1eq9.11}));\ 
broken lines: the low-energy expansion (\ref{1eq9.13}) up to order $k^2$;\ 
dashed lines: the high-energy expansion (\ref{1eq9.12}) up to order $1/k^N$ ($N=2$ and $6$). 
(Since $k$ is real, ^^ ^^ $N=2$" and ^^ ^^ $N=6$" are, respectively, in effect $N=1$ and $N=5$ 
for ${\rm Im}\, R_r$.)

\noindent
(b) \ (example~5) \ $V(x)=2 \log \cosh x$; $x_0=-1/2$. 
Solid lines: the exact $R_r$ (equation (\ref{1eq9.16})); \ 
broken lines: the low-energy expansion (\ref{1eq9.18}) up to order $k^N$ ($N=2$ and $4$);\ 
dashed lines: the high-energy expansion (\ref{1eq9.17}) up to order $1/k^N$ ($N=2$, $6$, and $14$).
Note that the exact $R_r$ is singular at $k=1$. 

\noindent
(c) \ (example~6)\  $V(x)=\sqrt{-x}$; $x_0=-1$. 
Solid lines: the exact $R_r$ (equation (\ref{1eq9.19}));\ 
broken lines: the low-energy expansion (\ref{1eq9.22}) up to order $k^2$;\ 
dashed lines: the high-energy expansion (\ref{1eq9.21}) up to order $1/k^2$.
}
\end{figure}

The low-energy expansion for this potential is correctly given by (\ref{1eq6.33}) with (\ref{1eq6.22}):
\begin{eqnarray}
\label{1eq9.13}
\fl
R_r= -\tanh \frac{e^x}{2} 
- \left({\rm sech\,}\frac{e^x}{2}\right)^2\,{\rm Shi\,}(e^x)\, ik
\nonumber \\
 - 2\left({\rm sech\,}\frac{e^x}{2}\right)^3
\left[
\int_0^{e^x}\frac{\cosh(y-\case{1}{2}e^x)}{y}{\rm Shi\,}(y)\, dy
\right](ik)^2
+ \cdots,
\end{eqnarray}
where ${\rm Shi\,}(z)=\int_0^z(1/t) \sinh t \,dt$ is the hyperbolic sine integral function (figure~2(a)).
The radius of convergence of (\ref{1eq9.13}) is larger than $1/2$, and 
it approaches $1/2$ as $x \to -\infty$. 

\medskip
\noindent
{\bf Example 5}\,: \quad $V(x)=2 \log \cosh x, \quad f(x)=-\tanh x.$

\nobreak
\medskip
\noindent
This is another example of a potential that grows linearly as $x\to -\infty$. 
The exact form of the reflection coefficient is expressed in terms of 
the hypergeometric function 
$
F(\alpha,\beta,\gamma;z)
=\frac{\Gamma(\gamma)}{\Gamma(\alpha)\Gamma(\beta)}
\sum_{n=0}^\infty
\frac{\Gamma(\alpha+n)\Gamma(\beta+n)}{\Gamma(\gamma+n)}
\frac{1}{n!}z^n
$.
We define
\begin{eqnarray}
\label{1eq9.14}
\fl
\eta(x,k)&\equiv 
F\left(\alpha, \beta, \case{1}{2}; -\sinh^2 x\right) 
+2 \frac{\Gamma\left(\frac{1}{2}+\alpha\right)\Gamma(1-\beta)}
{\Gamma(\alpha)\Gamma\left(\frac{1}{2}-\beta\right)}
\sinh x 
F\left(\alpha+\case{1}{2}, \beta+\case{1}{2}, \case{3}{2};-\sinh^2 x\right)
\nonumber
\end{eqnarray}
with
$\alpha \equiv \frac{1}{2}[-1-ik \sqrt{1-(1/k)^2}]$, 
$\beta \equiv \frac{1}{2}[-1+ik \sqrt{1-(1/k)^2}].$
Then we have
\begin{equation}
\label{1eq9.16}
R_r(x,-\infty;k)=\frac{ik\eta(x,k) + \eta'(x,k)}{ik\eta(x,k) - \eta'(x,k)},
\end{equation}
where $\eta'=\partial \eta/\partial x$.
The high-energy expansion (\ref{1eq7.20}) now takes the form
\begin{eqnarray}
\label{1eq9.17}
\fl
R_r &= \frac{\tanh x}{2ik}+\frac{({\rm sech\,}x)^2}{(2ik)^2}
-\frac{(3 + \cosh 2x) ({\rm sech \,}x)^2\tanh x}{2(2ik)^3}
-\frac{(3 + \cosh 2x)({\rm sech \,}x)^4}{2(2ik)^4}
\nonumber \\
\fl
& \qquad
+\frac{(11+ 8 \cosh 2x+ \cosh 4x) ({\rm sech \,}x)^4\tanh x}{4(2ik)^5}+\cdots.
\end{eqnarray}
As shown in figure~2(b), this is the correct large-$\vert k \vert$ expansion of (\ref{1eq9.16})  for ${\rm Im}\,k\ge 0$.
As in example~1, the series (\ref{1eq9.17}) is convergent for $\vert k \vert >1$. 

Calculating (\ref{1eq6.20}) for this potential, we obtain the low-energy expansion
\begin{eqnarray}
\label{1eq9.18}
\fl
R_r&= 1+2(\cosh x)e^x ik + 2(\cosh x)^2 e^{2x} (ik )^2 
+\left[2(\cosh x)^3e^{3x} -(\cosh x)^2e^{2x}\right] (ik)^3
\nonumber \\
\fl
& \qquad 
+\left[2(\cosh x)^4e^{4x} -4(\cosh x)^4e^{2x}+2(\cosh x)^3e^x\right](ik)^4
+ \cdots.
\end{eqnarray}
This is the correct small-$\vert k \vert$ expansion of (\ref{1eq9.16}), as can be seen from figure~2(b).

\medskip
\noindent
{\bf Example 6}\,: 
\quad $\displaystyle V(x)=\sqrt{-x}, \quad f(x)=\frac{1}{4}\frac{1}{\sqrt{-x}}.$
\quad ($x_{\rm max}=0$.)

\nobreak
\medskip
\noindent
In this example, $V(x)$ slowly tends to infinity as $x \to -\infty$, 
while $f(x)$ slowly converges to zero. 
The exact $R_r(x,-\infty)$ $(x<0)$ has the form
\begin{eqnarray}
\label{1eq9.19}
R_r(x,-\infty;k)=\frac{b(x,k)}{a(x,k)},
\\
\fl
a(x,k)\equiv
2\Gamma\left(\alpha+1\right)
F\left(\alpha,\case{1}{2};2ikx\right)
-\frac{1}{4}\left(\frac{i}{2k}\right)^{1/2}\,\Gamma\left(\alpha+\case{1}{2}\right)
\sqrt{-x}\,
F\left(\alpha+\case{1}{2},\case{3}{2};2ikx\right),
\nonumber \\
\fl
b(x,k)\equiv -\Gamma\left(\alpha+1\right)
\sqrt{-x}\,F\left(\alpha+1,\case{3}{2};2ikx\right)
+\frac{1}{2}\left(\frac{i}{2k}\right)^{1/2}
\Gamma\left(\alpha+\case{1}{2}\right)
F\left(\alpha+\case{1}{2},\case{1}{2};2ikx\right),
\nonumber
\end{eqnarray}
where $\alpha \equiv i/(32k)$. 
Expression (\ref{1eq7.20}) now reads
\begin{equation}
\label{1eq9.21}
\fl
R_r= -\frac{(-x)^{-1/2}}{8ik}-\frac{(-x)^{-3/2}}{8(2ik)^2}
-\frac{(x+12)(-x)^{-5/2}}{64(2ik)^3}
-\frac{(5x+60)(-x)^{-7/2}}{128(2ik)^4}
+ \cdots.
\end{equation}
This is the high-energy asymptotic expansion of (\ref{1eq9.19}) 
which is valid as long as $x < 0$. This expansion holds even for real $k$.
(See figure~2(c).)

The low-energy expansion is obtained from (\ref{1eq6.20}) as
\begin{eqnarray}
\label{1eq9.22}
\fl
R_r&=
1+4\left(1+\sqrt{-x}\,\right)ik + 8 \left(1+\sqrt{-x}\,\right)^2 (ik)^2
-32\left(5 + 4 \sqrt{-x} -x \right)(ik)^3
\nonumber \\
\fl
& \qquad
+32 \left(-21-40\sqrt{-x}+26x+8x\sqrt{-x}-x^2\right)(ik)^4 + \cdots.
\end{eqnarray}
This is the correct asymptotic expansion of (\ref{1eq9.19}) (figure~2(c)).
Unlike examples 1--5, this series is divergent;
the $R_r(k)$ given by (\ref{1eq9.19}) is essentially singular at $k=0$.

\medskip
\noindent
{\bf Example 7}\,: \quad $\displaystyle V(x)=\alpha \log (-x), 
\quad f(x)=-\frac{\alpha}{2x}.$
\quad ($x_{\rm max}=0$.)

\nobreak
\medskip
\noindent
Here $\alpha$ is a positive constant.
This potential tends to infinity 
as $x \to -\infty$ even more slowly than the previous one. 
The exact $R_r(x,-\infty)$ $(x<0)$ is expressed in terms of the Bessel function as
\begin{equation}
\label{1eq9.23}
\fl
R_r(x,-\infty;k)=
\frac{[J_\nu(-kx)+i J_{\nu-1}(-kx)]-i e^{i\alpha\pi/2}[J_{-\nu}(-kx)-i J_{1-\nu}(-kx)]}
{[J_\nu(-kx)-i J_{\nu-1}(-kx)]-i e^{i\alpha\pi/2}[J_{-\nu}(-kx)+i J_{1-\nu}(-kx)]},
\end{equation}
where $\nu \equiv (1+\alpha)/2$.
From (\ref{1eq7.20}) and (\ref{1eq7.17}) we have
\begin{equation}
\label{1eq9.25}
R_r= \frac{\alpha}{4xik}-\frac{\alpha}{2x^2(2ik)^2}
-\frac{\alpha^3-8 \alpha}{8 x^3 (2ik)^3}
+\frac{5\alpha^3-24\alpha}{8 x^4 (2ik)^4}
+\cdots.
\end{equation}
We can check that this high-energy expansion is correct even when ${\rm Im}\,k=0$.

On the other hand, the low-energy expression (\ref{1eq6.33}) for this $V(x)$ reads
\begin{equation}
\label{1eq9.26}
R_r= 1+2(-x)^\alpha ik\int_{-\infty}^x\frac{1}{(-z)^\alpha}\,dz+ \cdots,
\end{equation}
but the integral on the right-hand side is divergent if $\alpha\leq1$.
From the exact expression (\ref{1eq9.23}) we can see that the correct asymptotic form for $\alpha \leq1$ is
\begin{equation}
\label{1eq9.27}
R_r =1-2^{1-\alpha}
\frac{\Gamma\left(\frac{1-\alpha}{2}\right)}{\Gamma\left(\frac{1+\alpha}{2}\right)}
(-x)^\alpha k^\alpha
+\cdots,
\end{equation}
which includes a fractional power of $k$. If $\alpha>1$, then (\ref{1eq9.26}) is correct to order $k$, but the expansion in integral powers of $k$ fails at some higher order. 

\medskip
\noindent
{\bf Example 8}\,: {\it Potential with a singularity.}

\nobreak
\medskip
\noindent
As an example of a potential that has a singularity on the real axis, let us consider 
\begin{equation}
\label{2eq9.32}
V(x)=
\cases{
e^{-x} & $(x<0)$
\\
1-x & $(x>0)$
},
\qquad 
f(x)=
\cases{
\case{1}{2}e^{-x} & $(x<0)$
\\
\case{1}{2} & $(x>0)$
}.
\end{equation}
In this case, $f(x)$ is continuous and piecewise differentiable. 
The derivative of $f(x)$ has a jump at $x=0$. 
The exact $R_r(x,-\infty)$ for $x>0$ has the form
\begin{equation}
\label{2eq9.33}
\fl
R_r(x,-\infty;k)
=\frac{
-ik\left(A -2\right)B+1
+ \left[-ik\left(A +2\right)B-1\right]
e^{ikAx}
}
{
-ik\left(A +2\right)+B
+ \left[-ik\left(A-2\right)-B\right]
e^{ikAx}
},
\end{equation}
\begin{equation}
\label{2eq9.34}
\fl
A \equiv \sqrt{4-(1/k^2)},
\qquad
B \equiv 
i \frac{J_{-\nu}(-i/2)-ie^{-k\pi/2}J_{\nu}(-i/2)}
{J_{1-\nu}(-i/2)+ie^{-k\pi/2}J_{\nu-1}(-i/2)},
\qquad \nu \equiv ik+\frac{1}{2}.
\end{equation}
(This $B$ is the value of (\ref{1eq9.8}) at $x=0$.)  
Now (\ref{1eq7.20}) and (\ref{1eq7.17}) yield the expansion
\begin{equation}
\label{2eq9.36}
R_r= -\frac{1}{4ik}+ \frac{1}{(4ik)^3}-\frac{2}{(4ik)^5}+\frac{5}{(4ik)^7} +\cdots.
\end{equation}
If the limit $\vert k \vert \to \infty$ is taken with $0<\arg k <\pi$, then the $e^{ikAx}$ in (\ref{2eq9.33}) falls off faster than any power of $1/k$, and we can show that (\ref{2eq9.36}) is the correct asymptotic expansion.
(See the comments at the end of section~6.) 
If ${\rm Im}\,k=0$, then the $e^{ikAx}$ cannot be neglected. In this example, (5.15) does not hold for any $N$ when ${\rm Im}\,k=0$. 
The low-energy expression (\ref{1eq6.33}) is valid irrespective of the presence of the singularity.


\section{Summary}
The generalized reflection coefficient for the semi-infinite interval,
$\bar R_r(x,-\infty;\xi)$, 
is expressed in the form of (\ref{1eq5.19}) 
in terms of the operators ${\cal A}$ and ${\cal B}$ defined by (\ref{1eq5.1}) and (\ref{1eq5.2}).
Using the operator equations (\ref{3eq5.35a}) and (\ref{3eq5.35b}), 
with ${\cal L}$ defined by (\ref{1eq5.27}), we can derive expansions of $\bar R_r$ 
in powers of $k$ and $1/k$, together with the remainder terms (equations (\ref{3eq6.6}) and (\ref{1eq7.3})).
For either the low-energy or the high-energy expansion, the remainder term is expressed in terms of the inverse operator $({\cal A}-2ik{\cal B})^{-1}$, and, according to (\ref{1eq5.13}), it can be written as integrals involving the scattering coefficients ((\ref{1eq6.16}) and (\ref{1eq7.12})). 
By using the asymptotic forms of the scattering coefficients given in appendix~A, we can 
study the behavior of the remainder term as $k \to 0$ or $\vert k \vert \to \infty$, 
and investigate whether the expansion is asymptotic or not. 
The results are roughly summarized in figure~3.
For the high-energy expansion, conditions concerning differentiability of the potential must also be taken into account, as explained in section~6.
\begin{figure}
\hspace{1cm}
\includegraphics[scale=0.7]{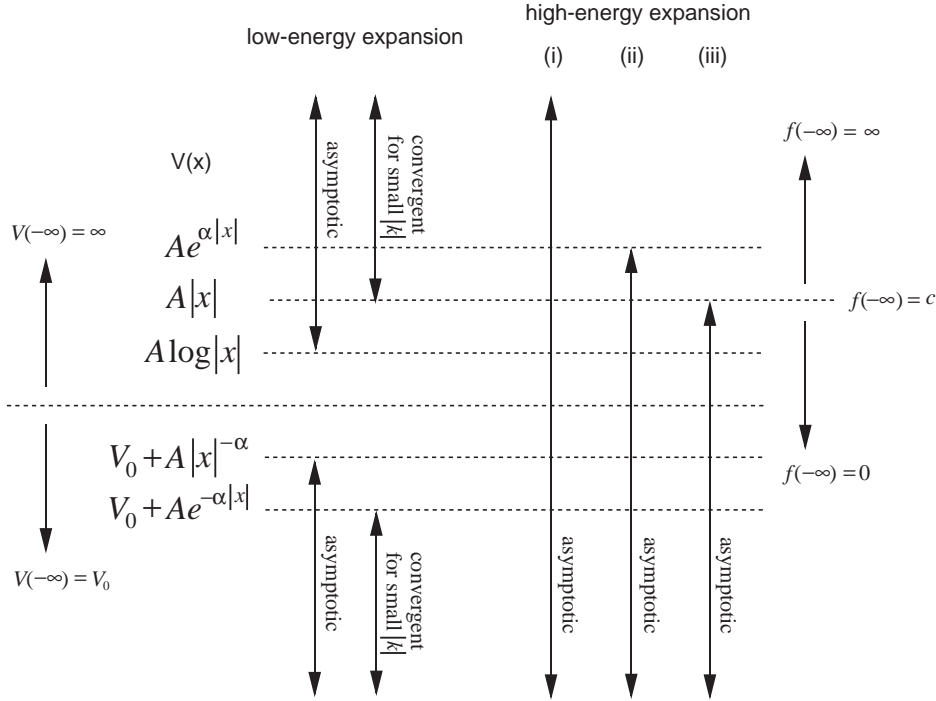}
\caption{Domains of validity of the low-energy expansion (\ref{1eq6.33}) and the high-energy expansion (\ref{1eq7.20}), 
where the limit $\vert k \vert \to \infty$ is taken with: (i) fixed $\arg k$ ($0<\arg k<\pi$); 
(ii) fixed ${\rm Im}\,k>0$; (iii) ${\rm Im}\,k=0$.}
\end{figure}
(The problem of whether the high-energy expansion is convergent or not is 
beyond the scope of this paper.)


\appendix


\section{Asymptotic behavior of $\bar \tau(x,y)$ and $\bar R_l(x,y)$ 
as $y \to -\infty$}
Here we summarize the asymptotic forms of $\bar \tau(x,y;\xi;k)$ and $\bar R_l(x,y;\xi;k)$ as $y \to -\infty$ 
with fixed $x$, $\xi$, and $k$ (${\rm Im}\,k\geq 0$).
(The derivation is omitted for space limitations.)

\bigskip
\noindent
(i) $f(-\infty)=\pm \infty$

\nobreak
\medskip
\noindent
In this case the asymptotic form of $\bar \tau(x,y)$ as $y\to -\infty$ is
\begin{equation}
\label{1eq3.1}
\bar \tau(x,y;\xi;k) = C(x,\xi,k) \exp \left[-\frac{1}{2}\vert V(y)\vert+ \eta(y,k)\right][1+o(1)],
\end{equation}
where $\eta(y,k)=o(\vert y \vert)$ $(y \to -\infty)$.
If $1/f(y)=O(1/\vert y \vert^{1+\epsilon})$ with some positive number $\epsilon$,  
then $\eta(y,k)=O(1)$. In that case we may let $\eta(-\infty,k)$ be absorbed into the $y$-independent quantity $C$, and redefine $\eta$ to be identically zero.
The $\bar R_l$ behaves as
\begin{equation}
\label{1eq3.3}
\bar R_l(x,y;\xi;k)= \mp1-\frac{ik}{f(y)}[1+o(1)],
\end{equation}
where the $\mp 1$ on the right-hand side corresponds to $f(-\infty)=\pm \infty$, respectively.
 
\bigskip
\noindent
(ii) $f(-\infty)=c\quad( c\neq 0, \pm\infty)$

\nobreak
\medskip
\noindent
When ${\rm Re\,}\sqrt{c^2-k^2}>0$ (i.e.,  ${\rm Im\,}k>0$ or ${\rm Im \,}k=0$, $c^2>k^2$),
we have, as $y \to -\infty$,
\begin{equation}
\label{1eq3.4}
\bar \tau(x,y;\xi;k)=C(x,\xi,k) \exp\left[\sqrt{c^2-k^2}\,y +\eta(y,k) \right] [1+o(1)],
\end{equation}
\begin{equation}
\label{1eq3.5}
\bar R_l(x,y;\xi;k)= -\frac{1}{c}\left(ik + \sqrt{c^2-k^2}\,\right)+o(1),
\end{equation}
where $\eta(y,k)=o(\vert y \vert)$.
If $f(y)=c+O(1/\vert y \vert^{1+\epsilon})$ with some positive $\epsilon$, then 
$\eta(y,k)=O(1)$, and so we may take $\eta$ to be zero. 

When $k$ is real and $k^2 \geq c^2$, 
equations (\ref{1eq3.4}) and (\ref{1eq3.5}) do not hold.
When evaluating an integral like (\ref{1eq5.13}) in such cases, we need to add an imaginary part $i\epsilon$ ($\epsilon>0$) to $k$, and let $\epsilon \to 0$ afterwards.
This is because $\bar R_r(k)$ has branch cuts along the real axis.
 
\bigskip
\noindent
(iii) $f(-\infty)=0$

\nobreak
\medskip
\noindent
In this case we have, as $y \to -\infty$,
\begin{equation}
\label{1eq3.8}
\bar \tau(x,y;\xi;k)=C(x,\xi,k) \exp[-iky +i\theta(y,k)][1+o(1)], \quad k \neq 0,
\end{equation}
where $\theta(y,k)=o(\vert y\vert)$.
If $f^2(y)=O(1/\vert y \vert^{1+\epsilon})$, then we may let $\theta=0$. 
If ${\rm Im}\,k>0$, then 
\begin{equation}
\label{1eq3.10}
\bar R_l(x,y;\xi;k)= o(1). 
\end{equation} 
If ${\rm Im}\,k=0$, then $\bar R_l(x,y)$ does not vanish but oscillates 
as $y \to -\infty$:
\begin{equation}
\label{1eq3.11}
\bar R_l(x,y;\xi;k)= D(x,\xi,k) \exp \left[-2ik y +2i\theta(y,k)\right]+o(1),\quad k \neq 0,
\end{equation}
where $D(x,\xi,k)$ is another quantity independent of $y$. 
The $\theta(y,k)$ in (\ref{1eq3.11}),  which is the same one as in (\ref{1eq3.8}), is a real quantity when $k$ is real.


\section{Proof of (\ref{1eq5.15}) with (\ref{1eq5.13})}

We are assuming that $g(x,\xi)$ is an analytic function of $\xi$ in $-1<\xi<1$, as mentioned at the beginning of section~5.
So we may expand $g/(1-\xi^2)$ in powers of $\xi$ and write 
\begin{equation}
\label{2eqc.1}
g(x,\xi)\equiv (1-\xi^2)\sum_{n=0}^\infty\xi^n h_n(x).
\end{equation}
We consider each term of (\ref{2eqc.1}) separately. 
Calculating with (\ref{1eq5.1}) and (\ref{1eq5.2}), we have
\begin{eqnarray}
\label{1eqc.1}
\fl
({\cal A}-2ik{\cal B})(1-\xi^2)\xi^nh_n(x)
\nonumber \\
\fl
\qquad =
(1-\xi^2) \Biggl\{
\xi^n\frac{dh_n(x)}{dx} +\left[n \xi^{n-1}-(n+2)\xi^{n+1}\right] f(x) h_n(x)
-2(n+1)ik \xi^nh_n(x) \Biggr\}.
\nonumber \\
\end{eqnarray}
Let us apply $({\cal A}-2ik {\cal B})^{-1}$ given by (\ref{1eq5.13}) to the first term on the right-hand side:
\begin{eqnarray}
\label{1eqc.2}
\fl
\frac{1}{{\cal A}-2ik {\cal B}}(1-\xi^2)\xi^n \frac{dh_n(x)}{dx}
=\int_{-\infty}^x \bar \tau^2(x,z;\xi)\bar R_l^n(x,z;\xi)\frac{dh_n(z)}{dz}\,dz
\nonumber \\
\fl
\qquad =\tau^2(x,z;\xi)\bar R_l^n(x,z;\xi)h_n(z)\Big\vert_{z=-\infty}^{z=x}
-\int_{-\infty}^x\frac{\partial}{\partial z}\left[\bar \tau^2(x,z;\xi)\bar R_l^n(x,z;\xi)\right]h_n(z)\,dz.
\end{eqnarray}
It can be shown that $\tau$ and $R_l$ satisfy the differential equations~\cite{algebraic}
\begin{eqnarray}
\label{1eqc.3}
\frac{\partial}{\partial z}\bar \tau(x,z)=-ik \bar\tau(x,z) - f(z) \bar \tau(x,z)\bar R_l(x,z), 
\\
\label{1eqc.4}
\frac{\partial}{\partial z}\bar R_l(x,z)=-2ik\bar R_l(x,z) + f(z) \left[1-\bar R_l^2(x,z)\right].
\end{eqnarray}
Using (\ref{1eqc.3}) and (\ref{1eqc.4}), we can rewrite the integral in the last expression of (\ref{1eqc.2}) as
\begin{eqnarray}
\label{1eqc.5}
\fl
&\int_{-\infty}^x\frac{\partial}{\partial z}\left[\bar \tau^2(x,z;\xi)\bar R_l^n(x,z;\xi)\right]h_n(z)\,dz
\nonumber \\
\fl
& \qquad = \int_{-\infty}^x \bar \tau^2\left[
n \bar R_l^{n-1}-(n+2)\bar R_l^{n+1}\right]fh_n  \,dz
 - 2 (n+1) ik \int_{-\infty}^x \bar \tau^2\bar R_l^n h_n \,dz
\nonumber \\
\fl
&\qquad =
\frac{1}{{\cal A}-2ik {\cal B}}(1-\xi^2)
\left\{\left[n \xi^{n-1}-(n+2)\xi^{n+1}\right] f(x) h_n(x)
-2(n+1)ik \xi^nh_n(x) \right\}.
\nonumber \\
\fl
\end{eqnarray}
From (\ref{1eqc.1}), (\ref{1eqc.2}), and (\ref{1eqc.5}) we have
\begin{equation}
\label{1eqc.6}
\fl
\frac{1}{{\cal A}-2ik {\cal B}}({\cal A}-2ik{\cal B})(1-\xi^2)\xi^nh_n(x)
=\bar \tau^2(x,z;\xi)\bar R_l^n(x,z;\xi)h_n(z)\Big\vert_{z=-\infty}^{z=x}.
\end{equation}
Taking the sum over $n$ and using definition (\ref{2eqc.1}) gives
\begin{equation}
\label{1eqc.7}
\fl
\frac{1}{{\cal A}-2ik {\cal B}}({\cal A}-2ik{\cal B})g(x,\xi)
=\left.\frac{\bar \tau^2(x,z;\xi)}{1-\bar R_l^2(x,z;\xi)}\, g(z,\bar R_l(x,z;\xi))\right\vert_{z=-\infty}^{z=x}.
\end{equation}
If (\ref{1eq5.16}) is satisfied, then we can easily see, by using (\ref{6eq4.2}), that the right-hand side of (\ref{1eqc.7}) is equal to $g(x,\xi)$. 
Thus $({\cal A}-2ik {\cal B})^{-1}({\cal A}-2ik{\cal B})g=g$ holds.


\section{Proof of (\ref{4eq5.24})}

With $g=\bar R_r+\xi$, 
the expression to the right of the limit symbol in (\ref{1eq5.16}) reads
\begin{equation*}
\fl
\frac{
\bar \tau^2(x,z;\xi)
}{
1-\bar R_l^2(x,z;\xi)
}
\left[
\frac{R_r(z,-\infty)-\bar R_l(x,z;\xi)}
{1-R_r(z,-\infty)\bar R_l(x,z;\xi)}
+\bar R_l(x,z;\xi)
\right]
=\frac{\bar \tau^2(x,z;\xi) R_r(z,-\infty)}
{1-R_r(z,-\infty)\bar R_l(x,z;\xi)}.
\nonumber
\end{equation*}
From (\ref{1eq3.1}), (\ref{1eq3.4}), and (\ref{1eq3.8}), we can see that $\lim_{z\to -\infty} \bar \tau(x,z)=0$ if $f(-\infty)=\pm \infty$ 
or ${\rm Im}\,k>0$. On the other hand, if $f(-\infty)=0$ then $\lim_{z\to -\infty}R_r(z,-\infty)=0$.
Therefore, the right-hand side vanishes in the limit $z \to -\infty$ for 
all $k$ with ${\rm Im}\,k\geq 0$, except in the case $f(-\infty)=c\, (\neq 0,\pm\infty)$ with ${\rm Im}\,k=0$. (For this exceptional case, see the comment in (ii) of appendix~A.)


\section{Explicit forms of (\ref{1eq6.19}) and the remainder terms}

The right-hand sides of (\ref{1eq6.19a}) and (\ref{1eq6.19b}) can be explicitly calculated. 
The result is
\begin{equation}
\label{6eqe.1}
C^\pm_{s_1,s_2,\ldots,s_{n-1}}(W)
=c^\pm_{s_1,\ldots,s_{n-1}}\exp\biggl[\Bigl(\pm 1-\sum_{i=1}^{n-1}s_i 
\Bigr)W\biggr],
\end{equation}
\begin{equation}
\label{b1eqd.2}
c^\pm_{s_1,\ldots,s_{n-1}} \equiv \pm 2 \prod_{j=1}^{n-1}\biggl[\mp s_j\Bigl(1\mp \sum_{k=1}^j s_k\Bigr)\biggr].
\end{equation}
The expression for $D_{s_1,s_2,\ldots,s_{n-1}}(W)$ can be written in the form
\begin{equation}
\label{3eqe.3}
\fl
D_{s_1,s_2,\ldots,s_{n-1}}(W)
={\rm sech}^{n+1} [(V_0-W)/2]
\sum_{m=-n+1}^{n-1} d_{s_1,\ldots,s_{n-1};m} \exp\left(\case{1}{2}mW\right).
\end{equation}
where $d_{s_1,\ldots,s_{n-1};m}$ are constants.
(We omit writing out the expressions for them.) 

Substituting (\ref{1eq6.18}) (with (\ref{6eqe.1}) and (\ref{3eqe.3})) into (\ref{1eq6.17}) and (\ref{1eq6.16}), we may write
\begin{eqnarray}
\label{1eq6.25}
\fl
\bar \rho_n&=(ik)^{n+1}\sum_{\{s_1,\ldots,s_n\}}
\int_{-\infty}^x\frac{\bar \tau^2(x,z)}{1-\bar R_l^2(x,z)}
\,P^+_{s_1,\ldots,s_n}(x,z)
\nonumber \\
\fl
& \qquad \qquad \qquad \qquad  
\times[-1,s_1,s_2,\ldots,s_{n-1}]_{-\infty}^z e^{s_nV(z)}\,dz,
\qquad V(-\infty)=+\infty,
\nonumber \\
\fl
&=(ik)^{n+1}\sum_{\{s_1,\ldots,s_n\}}
\int_{-\infty}^x\frac{\bar \tau^2(x,z)}{1-\bar R_l^2(x,z)}
\,P^-_{s_1,\ldots,s_n}(x,z)
\nonumber \\
\fl
& \qquad \qquad \qquad \qquad 
\times[+1,s_1,s_2,\ldots,s_{n-1}]_{-\infty}^z e^{s_nV(z)}\,dz,
\qquad V(-\infty)=-\infty,
\nonumber \\
\fl
&=(ik)^{n+1}\sum_{\{s_1,\ldots,s_n\}}
\int_{-\infty}^x\frac{\bar \tau^2(x,z)}{1-\bar R_l^2(x,z)}
\,Q_{s_1,\ldots,s_n}(x,z)
\nonumber \\
\fl
& \qquad \qquad \qquad \qquad  
\times(\pm,s_1,s_2,\ldots,s_{n-1}]_{-\infty}^z e^{s_nV(z)}\,dz,
\qquad V(-\infty)=V_0,
\end{eqnarray}
where
\begin{equation}
\label{4eqe.5}
\fl
P^\pm_{s_1,s_2,\ldots,s_n}(x,z)=
c^\pm_{s_1,\ldots,s_n}
\left[
\frac{1+\bar R_l(x,z)}{1-\bar R_l(x,z)}
\right]^{\pm 1-\sum_{i=1}^n s_i}
\exp\biggl[\Bigl(\pm 1-\sum_{i=1}^n s_i \Bigr)V(z)\biggr],
\end{equation}
\begin{eqnarray}
\label{3eqe.6}
\fl
Q_{s_1,s_2,\ldots,s_n}(x,z)
& =
\frac{\left[1-\bar R_l^2(x,z)\right]^{(n/2)+1}}
{\left\{
\cosh \case{1}{2}[V(z)-V_0]
+\bar R_l(x,z)\sinh \case{1}{2}[V(z)-V_0]
\right\}^{n+2}}
\nonumber \\
\fl
&\qquad \qquad \times
\sum_{m=-n}^n d_{s_1,\ldots,s_n;m}
\left[
\frac{1+\bar R_l(x,z)}{1-\bar R_l(x,z)}
\right]^{m/2}
\exp\left[\case{1}{2}mV(z)\right].
\end{eqnarray}


\section{Finiteness of $\bar \rho_N$ and $\bar \delta_N$}

The domain of $({\cal A}-2ik{\cal B})^{-1}$ is the range of ${\cal A}-2ik{\cal B}$ with its domain restricted to $\Omega^{[V]}_k$.
It is obvious that ${\cal A}\,\bar r_{N+1}$ belongs to the domain of $({\cal A}-2ik{\cal B})^{-1}$ if $\bar r_0 + \xi$ and $\bar r_1, \bar r_2, \ldots. \bar r_N$ all belong to $\Omega^{[V]}_k$. Namely, if $\bar r_0 + \xi$ and $\bar r_n$ ($1 \leq n \leq N$) satisfy condition (\ref{1eq5.16}), then (\ref{3eq6.8}) makes sense and is finite.
Using the asymptotic forms of $\bar \tau$ and $\bar R_l$ given in appendix~A, and the expressions for $\bar r_n$ given by (\ref{1eq6.18}) with (\ref{6eqe.1}) and (\ref{3eqe.3}), we can show that these conditions are satisfied as long as the $\bar r_n$ are finite.
(When $k$ is a nonzero real number, we need to be careful in the following two cases:
the case $f(-\infty)=c$ with $k^2>c^2\neq 0$, and the case where $f(-\infty)=0$ and $V(-\infty)=\pm \infty$. In these cases, the value of the integral (\ref{1eq6.16}) is indeterminate. So we need to replace $k$ by $k+i\epsilon$ with positive infinitesimal $\epsilon$, and let $\epsilon \to 0$ after evaluating the integral. Then the integral takas a definite value, and expression (\ref{3eq6.6}) is well-defined.)

Similarly, ${\cal B}\,\bar c_{N+1}$ lies in the domain of $({\cal A}-2ik{\cal B})^{-1}$ if $\bar c_1, \bar c_2, \ldots, \bar c_N$ belong to $\Omega^{[V]}_k$.
It is easy to see that this condition is satisfied as long as $\bar c_1(x,\xi), \ldots, \bar c_N(x,\xi)$ are continuous and piecewise differentiable with respect to $x$


\section{Verification of (\ref{1eq6.39})}

Here we study the cases $V(-\infty)=+\infty$ and $V(-\infty)=V_0$. The former is divided into two subcases, $f(-\infty)\neq 0$ and $f(-\infty)=0$. We omit the case $V(-\infty)=-\infty$, which is essentially the same as the case $V(-\infty)=+\infty$.

\bigskip
\noindent
(i)-A \quad $V(-\infty)=+\infty$, \quad $f(-\infty)\neq 0$ 

\nobreak
\medskip
\noindent
The first equation of (\ref{1eq6.25}) has the form
\begin{equation}
\label{1eqf.1}
\bar \rho_n
=
(ik)^{n+1}
\int_{-\infty}^x B_n(z,k)\bar \tau^2(x,z;k)\,dz.
\end{equation}
(In this appendix, we regard $x$ and $\xi$ as fixed constants.)
Using (\ref{1eq3.3}), (\ref{4eqe.5}), and (\ref{b1eqd.2}), we can show that $B_n(z,k)$ tends to a finite value as $z \to -\infty$. 

When $f(-\infty)=+\infty$ or $f(-\infty)=c\neq 0$, the behavior or $\bar \tau(x,z)$ 
as $z \to -\infty$ is given by (\ref{1eq3.1}) or (\ref{1eq3.4}). 
In either case there exist real constants $C_1$, $C_2$, and $k_1$ such that
$\vert \bar \tau^2(x,z;k) \vert 
\leq C_1 e^{-V(z)+C_2 z}$ 
for $z<x$ and $\vert k \vert < k_1$. 
Using this, we can easily show
\begin{equation}
\label{1eqf.5}
\fl
\lim_{k \to 0}\int_{-\infty}^x B_n(z,k)\bar \tau^2(x,z;k)\,dz
=\int_{-\infty}^x \lim_{k \to 0} B_n(z,k)\bar \tau^2(x,z;k)\,dz,
\end{equation}
which is equivalent to the second equality of (\ref{1eq6.39}).

\bigskip
\noindent
(i)-B \quad $V(-\infty)=+\infty$, \quad $f(-\infty) = 0$ 

\nobreak
\medskip
\noindent
If $V(x)$ tends to infinity 
more slowly than $\vert x \vert$ as $x \to -\infty$, the asymptotic behavior or $\bar \tau(x,z)$ is given by (\ref{1eq3.8}). The quantity $B_n(z,k)$ of (\ref{1eqf.1}) becomes infinite as $z \to -\infty$ for $n \geq 1$. 
We can show that $B_n$ does not grow faster than $\vert z \vert^n$, i.e., $B_n(z,k) =o(\vert z \vert^n)$.

Here we give only a sketchy explanation. Let $f(z)$ be monotone for $z<z_1$. For any $k$ satisfying $\vert k \vert < \vert f(z_1) \vert$, 
there exists a value $z_a(k)$ $(<z_1)$ such that $\vert k \vert=\vert f(z_a)\vert$.
As $k$ approaches zero, this $z_a$ tends to $-\infty$.
Let $k$ be sufficiently small. 
Then $\bar \tau(x,z;k)\simeq \bar \tau(x,z;k=0)$ for $z \gg z_a$,
and $\bar \tau(x,z;k) \simeq \bar \tau(x,z_a;k=0) \exp[-ik(z-z_a) + \cdots]$ for $z \ll z_a$.
Using $B_n(z,k) =o(\vert z \vert^n)$, we have the estimate
\begin{equation}
\label{1eqf.9}
\fl
\lim_{k\to 0} \left\vert \int_{-\infty}^{z_a(k)} B_n(z,k) \bar \tau^2(x,z;k) \,dz \right\vert
< C\, \lim_{k\to 0} \frac{e^{-2V(z_a(k))}}{k^{n+1}}
=C\lim_{z_a \to -\infty}\frac{e^{-2V(z_a)}}{f^{n+1}(z_a)},
\end{equation}
where $C$ is a constant. The last expression of (\ref{1eqf.9}) vanishes if
$V(z)$ tends to infinity 
faster than $\log \vert z \vert$ as $z \to -\infty$. 
Hence we can see that (\ref{1eqf.5}) holds even in this case.

\bigskip
\noindent
(ii) \quad $V(-\infty)=V_0 \neq \infty$

\medskip
\noindent
\nobreak
Let us consider each integral in the last expression of (\ref{1eq6.25}).
From (\ref{1eq3.8}), (\ref{1eq3.10}), and (\ref{3eqe.6}), we can see that there exist constants $C$ and $k_1$ such that
\begin{equation}
\label{1eqf.11}
\left\vert
\frac{\bar \tau^2(x,z;k)}{1-\bar R_l^2(x,z;k)}
\,Q_{s_1,\ldots,s_n}(x,z;k)
\right\vert
\le
C
\end{equation}
for any $z<x$ and $\vert k \vert< k_1$.
Thus, the integrand in the last expression of (\ref{1eq6.25}) is absolutely dominated by $C (\pm,s_1,s_2,\ldots,s_{n-1}]_{-\infty}^z e^{s_nV(z)}$, which is a $k$-independent function of $z$. The integral of this function is 
 $C\,(\pm,s_1,s_2,\ldots,s_{n-1},s_n]_{-\infty}^x$, which is finite if the potential satisfies condition (\ref{1eq6.24}).
Therefore we can interchange the limit $k \to 0$ and the integral in (\ref{1eq6.25}), and so the second equality of (\ref{1eq6.39}) holds.



\end{document}